\documentclass[aps,
amsmath,amssymb,
reprint,%
]{revtex4-2}
\usepackage{graphicx}
\usepackage{dcolumn}
\usepackage{bm}

\usepackage[utf8]{inputenc}
\usepackage[T1]{fontenc}
\usepackage{mathptmx}
\usepackage{etoolbox}

\makeatletter
\def\@email#1#2{%
	\endgroup
	\patchcmd{\titleblock@produce}
	{\frontmatter@RRAPformat}
	{\frontmatter@RRAPformat{\produce@RRAP{*#1\href{mailto:#2}{#2}}}\frontmatter@RRAPformat}
	{}{}
}%
\makeatother
\begin{document}
	

\title[]{Quasi-static remanence as a generic-feature of spin-canting in  Dzyaloshinskii-Moriya Interaction driven canted-antiferromagnets.}

\author{Namrata Pattanayak$^{1,2}$, Arun Kumar$^{1}$, A.K Nigam$^{3}$, Vladimir Pomjakushin$^{4}$, Sunil Nair$^{1}$, Ashna Bajpai$^{1}$}

\affiliation{$^1$Department of Physics, Indian Institute of Science Education and Research, Pune, India}

\affiliation{$^2$Department of Physics, NIST, Berhampur, Odisha, 761008, India}

\affiliation{$^3$Tata Institute of Fundamental research, India}

\affiliation{$^4$Laboratory for Neutron Scattering and Imaging, Paul Scherrer Institut, Villigen, CH-5232, Switzerland}

\email{ashna@iiserpune.ac.in}

\date{\today}
\begin{abstract}
		We  consistently observe a unique pattern in remanence in a number of canted-antiferromagnets (AFM) and piezomagnets. A part of the remanence  is \textit{quasi-static} in nature and vanishes above a critical magnetic field. Present work is devoted to  exploring this \textit{quasi-static} remanence ($\mu$) in a  series of isostructural canted-AFMs and piezomagnets that possess progressively  increasing  N\'eel temperature ($T{_N}$).  Comprehensive investigation of remanence  as a function of \textit{magnetic-field} and \textit{time}  in  CoCO$_{3}$, NiCO$_{3}$ and MnCO$_{3}$  reveals that the  magnitude of $\mu$  increases with decreasing $T{_N}$, but the stability with time is higher in the samples with higher $T{_N}$.  Further to this, all three carbonates exhibit a universal scaling in $\mu$, which relates to the concurrent phenomenon of piezomagnetism.   Overall, these data not only  establish that the observation of \textit{quasi-static} remanence with \textit{counter-intuitive} magnetic-field dependence can serve as  a foot-print for spin-canted systems, but also confirms that simple remanence measurements, using SQUID magnetometry, can provide insights about the extent of spin canting - a non trivial parameter to determine. In addition, these data suggest that the  functional form of $\mu$  with  \textit{magnetic-field} and \textit{time}  may hold key to isolate Dzyaloshinskii Moriya Interaction  driven spin-canted systems from Single Ion Anisotropy driven ones. We also demonstrate the existence of $\mu$ by tracking specific peaks in neutron diffraction data, acquired in remnant state  in CoCO$_{3}$.
\end{abstract}
	
\maketitle
	
Canted-antiferromagnets or \textit{weak ferromagnets} leading to exotic chiral spin structures form one of the most niche area of fundamental and applied research. \cite{Binz, Gayles, Jairo, Baltz, Ajejas, Di, Nvemec, Nicola}.The phenomenon of spin-canting is understood to arise from two primary mechanisms, one of which is the celebrated Dzyaloshinski -Moriya Intercation  (DMI)  and the other relates to Single Ion Anisotropy (SIA)\cite{Dzy1, Moriya1}. Hematite ($\alpha$-Fe$_2$O$_3$) along with a series of isostructural carbonates (MnCO$_{3}$, NiCO$_{3}$ \&  CoCO$_{3}$) and rutile fluorides (NiF$_{2}$ and CoF$_{2}$) constitute the canonical examples of DMI and SIA driven canted-AFMs respectively\cite{Dzy1, Moriya1}.  It is to be noted that spin canting in  CoCO$_{3}$  is understood to have contributions from both DMI as well as SIA \cite{Pincini}. This sets CoCO$_{3}$ apart from primarily DMI-driven canted AFM ($\alpha$-Fe$_2$O$_3$, MnCO$_{3}$ \&  NiCO$_{3}$)  and also from primarily SIA-driven canted AFM  (NiF$_{2}$ and CoF$_{2}$). All these canted AFMs, driven by either DMI or SIA are also known to be symmetry allowed piezomagnets (PzMs)\cite{Dzy1, Moriya1}.

Notably, some of these symmetry allowed canted-AFM \& PzMs have shown some unique features in \textit{remanence} measurements.  These measurements conducted on  single crystal, mesoscopic  \& nano-scale Hematite crystallites \cite{Pattanayak, Pattanayak1},  hematite inside carbon nanotubes \cite{AK1} as well as on the mesoscopic crystallites of MnCO$_3$  \cite{Pattanayak1} bring out two unique features. First is that there exist two  distinct time scales related to the magnetization dynamics, one of which is \textit{ultra-slow} in nature. This ultra slow magnetization relaxation results in  the observation of \textit{quasi static remanence},  which is referred to as $\mu$ throughout the text.  Second  feature is that the  $\mu$ shows a counter-intuitive magnetic field dependence.  These specific traits  set remanence in \textit{canted-AFM} apart from the one observed in  a routine  ferromagnetic (FM), antiferromagnetic or a  spin glass phase. The magnitude of this $\mu$ is tunable with  nanoscaling \cite{Pattanayak, Pattanayak1} as well as with  interface effects\cite{AK1} in above mentioned systems -which are symmetry allowed  DMI driven canted AFMs.  However,  Ultra thin films of Cr$_2$O$_3$ \cite{Binek}, composites of CrO$_2$/Cr$_2$O$_3$  \cite{Ashna1}  and encapsulation of Cr$_2$O$_3$ inside carbon nanotubes \cite{Ashna2} also exhibited very similar traits in $\mu$, implying novel strain \& interface effects leading to spin-canting and PzM.  

In the present work, we have extended our investigation pertaining to $\mu$  in a series of canonical spin-canted systems, that possess systematically  varying  $T{_N}$.  It is to be noted that larger spin canting angle is expected in samples with lower $T{_N}$. Also, the spin canting angle is a non trivial parameter to measure and it takes painstaking efforts to estimate the direction or the magnitude of spin canting\cite{Gross, Pincini, Beutier}. In this work, our endeavor is to establish the presence of $\mu$ as a  generic feature in spin-canted AFM and PzM and  explore  the  correlations  between this unique $\mu$  and the extent of spin-canting, which in -turn relates to $T{_N}$.  In addition, we  observe  a universal scaling in $\mu$ which has been earlier reported for symmetry allowed  piezomagnets, or samples in which piezomagnetism emerges due to size and interface effects \cite{Ashna1, Ashna2, Binek, Kleemann}.  Apart from remanence measurements using SQUID magnetometry, we also present a microscopic experimental evidence for existence of $\mu$ using  neutron diffraction experiments in a representative carbonate, CoCO$_3$.  

\section{Experimental Techniques}

\begin{figure}[!t]
\centering
\includegraphics[width=0.5\textwidth]{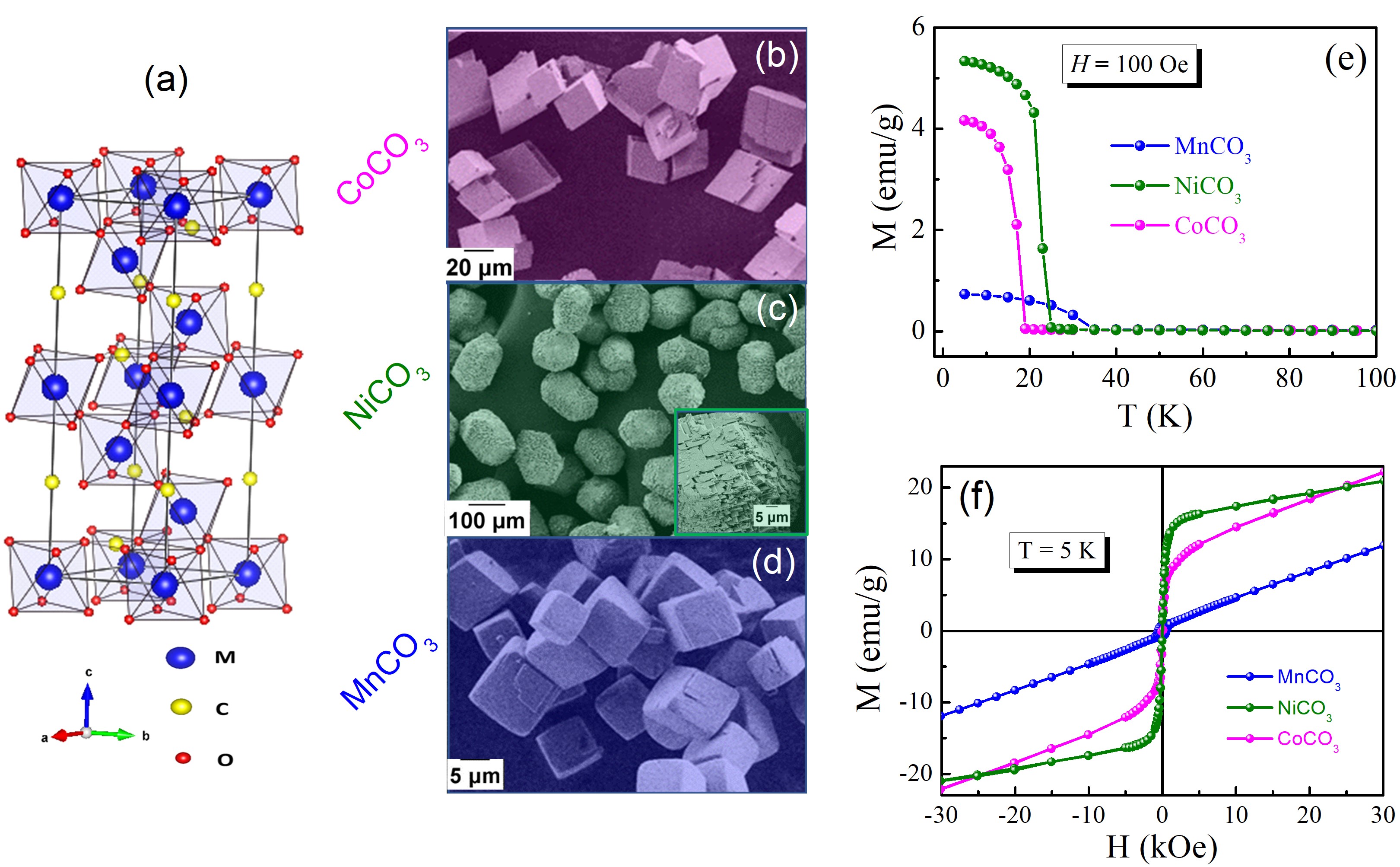}
\caption { \textbf{(a)}  Schematic  unit cell of isostructural carbonates with M being metal ion. \textbf{(b)}- \textbf{(c)} depict  SEM images of  CoCO$_{3}$, NiCO$_{3}$ and  MnCO$_{3}$ respectively. Samples consists of regular shaped polyhedra,  with size  ranging  from a few tens of microns to 100  microns. \textbf{(e)}  shows \textit{M} as a function of  \textit{T} measured at 100 Oe \& \textbf{(f)} shows \textit{M}  \textit{H} isotherms at 5K  for CoCO$_{3}$ (pink dots), NiCO$_{3}$ (green dots) and  MnCO$_{3}$ (blue dots).}
\label{Figure1}
\end{figure}

The carbonates presented in this work  are synthesized by the hydrothermal technique (Supp.Info: Text S1)  and the phase purity is confirmed using  X -ray diffraction, along with the Rietveld Profile refinement of the XRD data (Supp.Info : Figure S1). The lattice parameters determined from Rietveld profile refinement are shown in Table 1 in Supp. Information.  The schematic unit cell for the  carbonates, which crystallize in rhombohedral structure is depicted in  Figure~\ref{Figure1}(a) and the  scanning electron micrograph (SEM) are shown in  Figure~\ref{Figure1}(b)-(d) . The size range of individual crystallites falling in bulk limit ensures that the remanence data presented in this work is not associated with nano-scaling, which  can also lead to a slow-magnetization dynamics in routine ferromagnets \cite{Binder}.  

The \textit{M} versus \textit{T} are shown in Figure~\ref{Figure1}(e)  for all three carbonates.  The progressively increasing  $T{_N}$  values are 18K, 24K and 35 K for CoCO$_3$, NiCO$_3$ and  MnCO$_3$ respectively.  These values in good agreement with the previous reports \cite{ Kreines, Srivastava}. Figure~\ref{Figure1}(f)  compares  the \textit{MH} isotherms  at 5K for all three samples. Here  the M increases with increasing H and shows a slight opening of the loop in lower magnetic fields, followed by a  non -saturating behavior in higher fields. The loop opening in lower field region is also shown as Supp.Info, Figure S2.  This small  loop opening is a signature  of  weak ferromagnetism in otherwise AFMs. However, on a general note, it is  rather difficult to distinguish a canted AFM from a normal AFM or soft FM, a FM/AFM interface, or a glassy system based solely on magnetization measurement. These  systems  can also  show small opening of the loop in MH and mimic FM like behavior  M vs H or  M vs T. Microscopic measurements like neutron diffraction are therefore essential  for firmly establishing the AFM phase. It is all the more difficult to identify and  establish canting phenomenon  for samples in which the \textit{spin-canting} is arising due to size or interface effects. Our endeavor in the present work is to show that simple \textit{remanent magnetization} measurements can  better establish  a  canted AFM phase.  

\section{Results \& Discussion}

\subsection{\label{sec:level}  Temperature variation of Magnetization and the corresponding Remanence  in NiCO$_3$  }

\begin{figure*}[!t]
\centering
\includegraphics[width=0.9\textwidth]{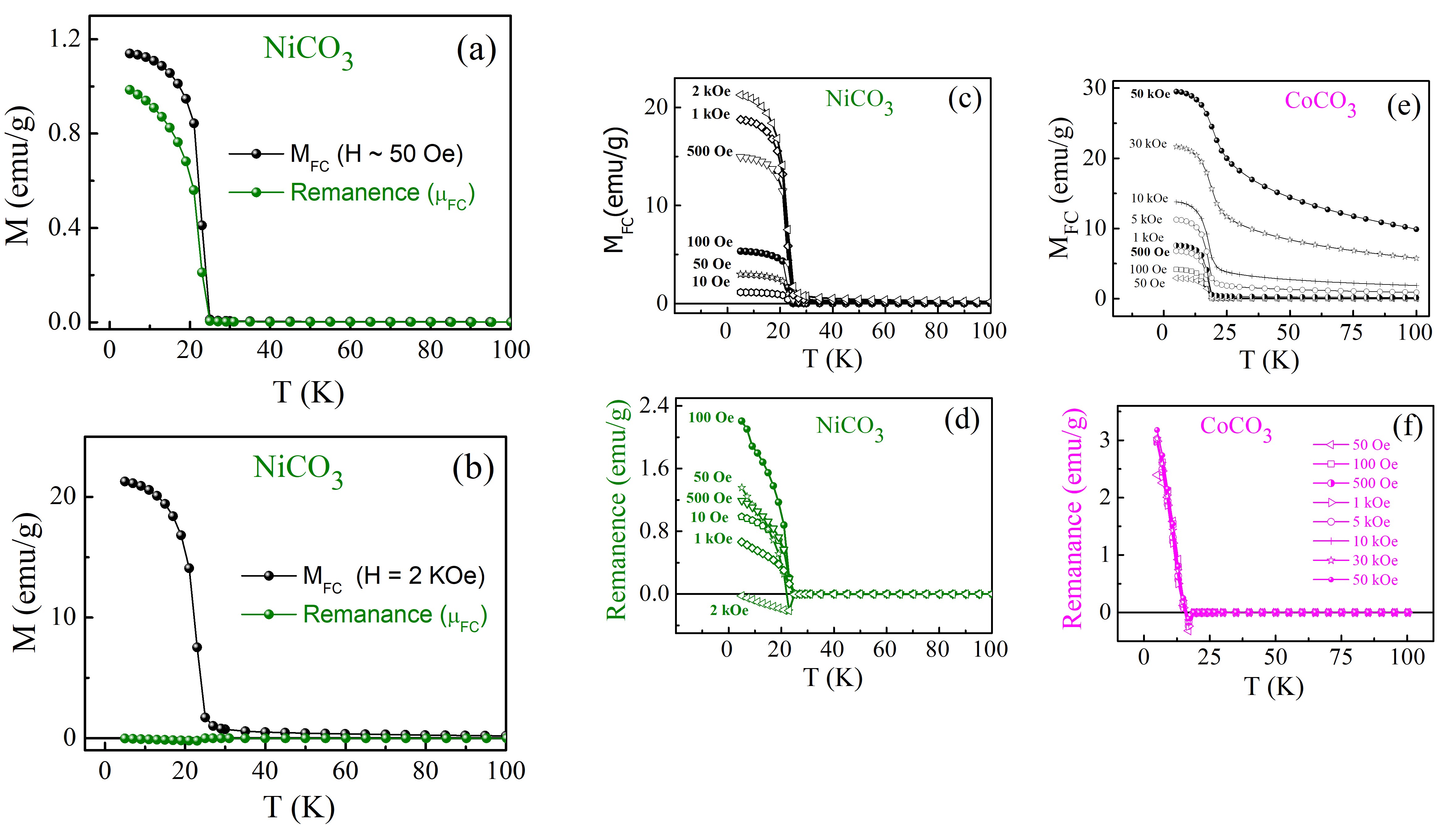}
\caption {\textbf{(a)} shows  \textit{M} versus \textit{T}  for NiCO$_3$  in presence of 100 Oe  while cooling the sample from above its  $T{_N}$ (black dots). At 5K the magnetic field is switched off and  corresponding remanence ($\mu$ versus \textit{T})  is measured in warming cycle (green dots).  \textbf{(b)} shows the same for  NiCO$_3$ in H =2 kOe.  \textbf{(c)}  shows \textit{M} versus \textit{T}(black dots) in various \textit{H} for NiCO$_3$. The  corresponding  remanence ($\mu$) versus \textit{T} (green dots) is plotted separately  in \textbf{(d)}. \textbf{(e)} \& \textbf{(f)} show  M and corresponding $\mu$ for for CoCO$_{3}$ in various (cooling) magnetic fields. }
\label{Figure2}
\end{figure*}

We first define the  protocol of magnetization and corresponding remanence measurements adopted in this work.  As shown in Figure~\ref{Figure2}(a),  the magnetization is measured while cooling the sample, which is  NiCO$_3$ in this case, from above its $T{_N}$. This is regular field-cooled cycle to obtain  $M{_{FC}}$ vs T data, shown as black dots in Figure~\ref{Figure2}(a). After reaching 5K, the applied \textit{H} is turned off and the magnetization  instantaneously drops. However it arrives to a fixed value, and does not show any further decay. This is basically the \textit{quasi-static} part of the remanence ($\mu$). We  note that $M{_{FC}}$ $\sim$ 1.14 emu/g and the corresponding $\mu$ $\sim$ 0.98 emu/g at 5K. Thus  the magnetization drops to about 95\% of its \textit{in-field} value.   It is to be emphasized that the  $\mu$  does not exhibit any further decay with time, as long as T is held at 5K.   On increasing the temperature, the  $\mu$ vs T can be measured in warming cycle (green dots in Figure~\ref{Figure2}(a)). It is evident that  the $\mu$ decreases with increasing T in a fashion which is qualitatively similar to $M{_{FC}}$ vs T. It vanishes in the paramagnetic region,  above the \textit{T}$_N$.  

	The magnitude of this $\mu$ depends on the magnetic field applied during the cooling cycle i.e, while measuring $M{_{FC}}$ vs T. For instance, when the  entire  protocol is repeated  for a larger cooling  \textit{H} of 2 kOe, the magnitude of \textit{M} at 5K is $\sim$ 21 emu/g and that of  the corresponding  $\mu$ is vanishingly small, $\sim$ 10$^{-4}$ emu/g , as is evident from Figure~\ref{Figure2}(b).  Thus, we observe that there is practically  no $\mu$ for NiCO$_3$ in magnetic field of 2kOe. The small negative value of $\mu$  for H ~2KOe case is an artifact of measurement. It arises  from the residual negative field of the superconducting magnet of SQUID. 

For all the intermediate (cooling) magnetic fields, \textit{M} vs T  and  the corresponding  $\mu$ vs T  is presented  in Figure~\ref{Figure2}(c) \& Figure~\ref{Figure2}(d)  respectively. The \textit{M} vs T and $\mu$ vs T are plotted  in separate graphs for the sake of clarity.  It is to be emphasized that all the remanence data  is obtained in zero H. The magnetic field specified in the Figure~\ref{Figure2}(d) refers to the  magnetic field used for preparing a remanent state while recording $M{_{FC}}$ vs T cycles. We note that  $\mu$ is nearly of same magnitude as the \textit{in-field} magnetization for lower (cooling) fields. The magnitude of $\mu$ is maximum for the (cooling) field of 100 Oe in case of NiCO$_3$, as is evident from Figure~\ref{Figure2}(d).  Also,  there is no \textit{quasi-static} $\mu$ above a critical magnetic field, which is  $\sim$ 2kOe in case of NiCO$_3$ (Figure~\ref{Figure2}(b)). For all the intermediate magnetic fields, the \textit{quasi-static} remanence first increases with increasing magnetic field up to  a critical field. Thereafter it sstarts to decreasing with increasing (cooling) field and eventually vanishes, as is evident from Figure~\ref{Figure2}(d). 

Following similar experimental protocol for preparing a remanent state in different magnetic fields, \textit{quasi-static} remanence has already been reported \cite{Pattanayak} for  MnCO$_3$ crystallites shown in Figure~\ref{Figure1}(d). This sample also shows a peak like behavior in $\mu$ as a function of magnetic field, before vanishing at another critical field \cite{Pattanayak}.     

\subsection{\label{sec:level} Temperature variation of Magnetization and the corresponding Remanence in CoCO$_3$  }

The M vs T  and the corresponding $\mu$ vs T  data for CoCO$_3$, following the similar experimental protocol as adopted for NiCO$_3$ is presented in  Figure~\ref{Figure2}(e)and Figure~\ref{Figure2}(f) respectively. CoCO$_3$ exhibits some  additional noteworthy features in M vs T cycles. For instance,  the transition region is fairly broadened, especially  when M is measured in higher magnetic fields. Broadening in the vicinity of magnetic  transition temperature  is  typically attributed to short range correlations.  However, in this series of carbonates, this broadening is substantial for CoCO$_3$  as compared to the other two carbonates and this feature extends much above the  T$_{N}$ of CoCO$_3$.   Another interesting observation that the $\mu$ vs T data nearly self -scale in case of CoCO$_3$ (Figure~\ref{Figure2}(f)). This pattern in $\mu$ vs T of CoCO$_3$ is strikingly different from both NiCO$_3$, shown in Figure~\ref{Figure2}(d) and MnCO$_3$ \cite{Pattanayak1}. Both these samples  also exhibit universal scaling,  albeit after the $\mu$ at a given temperature is normalized with $\mu$  measured at 5K. The issues of scaling will be discussed in section E.  It is also interesting to note that the $\mu$  is practically zero in the paramagnetic region for all three carbonates. We also note that the $\mu$ vanishes in the vicinity of \textit{T}$_N$ for each compound as is evident from Figure~\ref{Figure2} for NiCO$_3$ and  CoCO$_3$.  Qualitatively similar  data for MnCO$_3$ is reported in reference \cite{Pattanayak1}.  Thus we note that  $\mu$ vs T, especially in lower H, clearly marks the intrinsic AFM  transition,  better than the corresponding M vs T data in all three carbonates. In addition $\mu$ vs T measured at different H also reveal that  $\mu$ shows a peak like behavior and  vanishes above a critical field for both MnCO3 and NiCO3. This unique pattern of remanence as a function of magnetic field sets apart canted-AFM from other magnets, as shown in the next subsection.     

\subsection{\label{sec:level} Counter-intuitive magnetic field dependence of \textit{quasi static} remanence }

\begin{figure}[!t]
\centering
\includegraphics[width=0.4\textwidth]{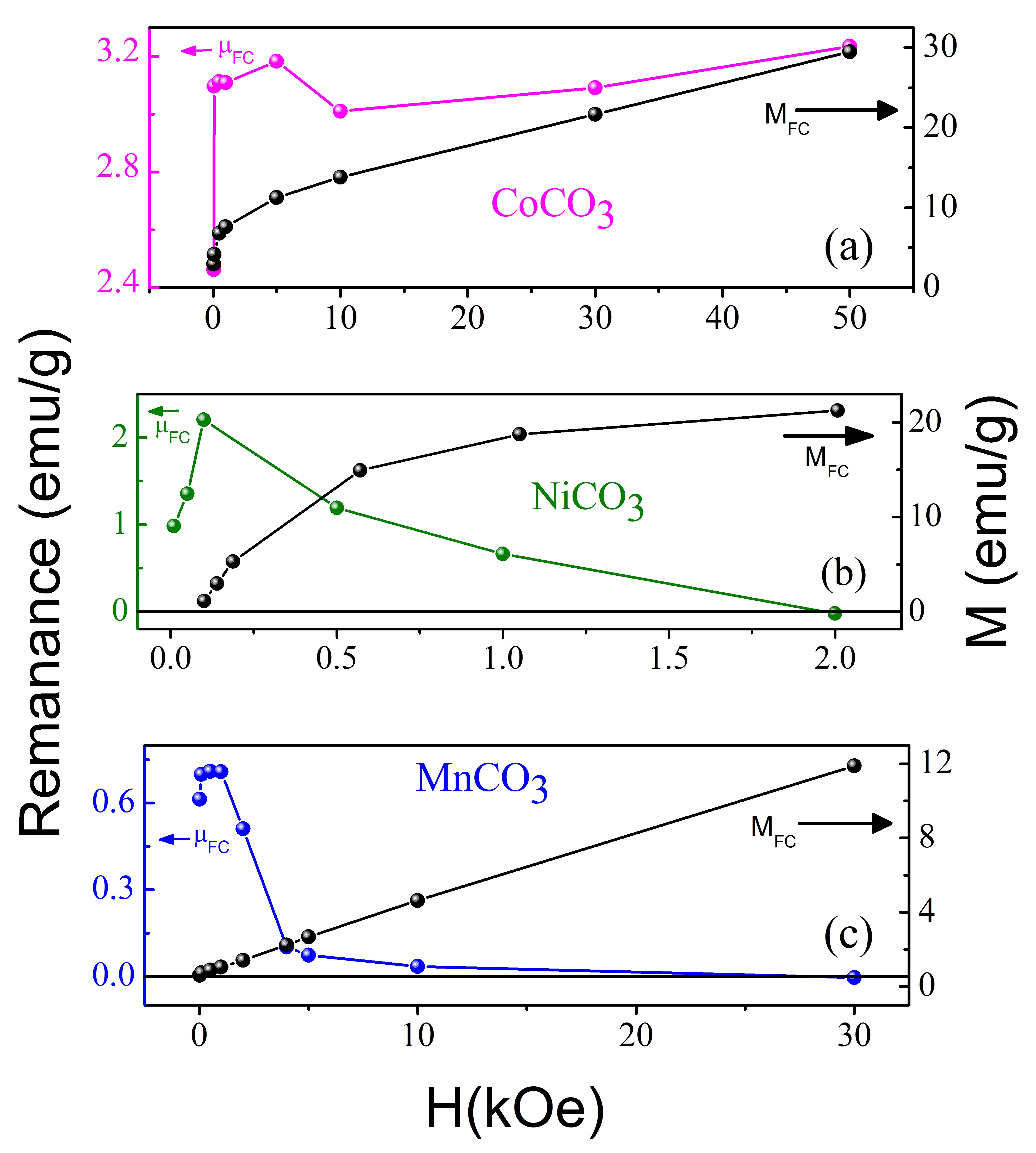}
\caption {\textbf{(a)}-\textbf{(c)}  show  field dependence of remanence ($\mu$ vs H)  for  all three carbonates in conjunction with the corresponding  M vs H (black dots) at a fixed temperature of 5K.  While M increases with increasing H, the corresponding remanence  exhibits  a peak like behaviour for MnCO$_{3}$  as well as NiCO$_{3}$. In case of  CoCO$_{3}$, which is understood to have contributions from SIA apart from DMI in spin-canting,  a tendency of peak-like behaviors is followed by  a rise in remanence upon  increasing H. }
\label{Figure3}
\end{figure}

Our core observation here is the counter-intuitive H dependence of $\mu$ in canted AFM, which is not reflected in the routine \textit{in-field} measurements \cite{Ashna1, Ashna2, Pattanayak, Pattanayak1}. To highlight this, we plot both M vs H in conjuncture with $\mu$ vs H  for all three carbonates at a fixed temperature of 5K.  Here M increases with increasing magnetic field and hence the field dependence of M  is regular for all three carbonates (black dots -right axis in Figure~\ref{Figure3}).  The corresponding $\mu$ (at 5K) shows an anomalous  H dependence for MnCO$_3$ (blue dots- left axis) and  NiCO$_3$ (green dots -left axis). In both cases, the magnitude of \textit{quasi-static} remanence  first increases with increasing H, followed by a drop. The remanence eventually vanishes at some H, depending on the sample. In case of NiCO$_3$  the critical magnetic field,  at which  the magnitude of $\mu$  is highest  is  100 Oe , as is evident from Figure~\ref{Figure3}(b). Another critical magnetic field at which the remanace vanishes  is about 2kOe for NiCO$_3$, as shown in Figure~\ref{Figure3}(b).  These critical fields  are  500 Oe and  30 kOe  respectively for MnCO$_3$, as shown in  Figure~\ref{Figure3}(c).  CoCO$_3$ also shows a tendency to peak like behavior, followed by a increases with increasing (cooling) magnetic field . It is to be noted that in case of CoCO$_3$, the $\mu$ does not vanish as shown in Figure~\ref{Figure3}(a).  It is to be noted that the magnitude of  $\mu$ is highest for  CoCO$_3$ in this series, which possesses the lowest $T{_N}$. 
	
	As we mentioned, the  counter-intuitive H dependence of the \textit{quasi-static} remanence   has been observed for a number of DMI driven canted AFM \cite{Ashna1, Ashna2, Pattanayak, Pattanayak1, AK1}. Associated physical mechanism appears to be related to the interplay between the exchange, Zeeman and the magnetocrystalline anisotropy,  with an additional factor  associated with the spin-canting. This in turn relates to a specific domain pattern \cite{AK1}. This domain pattern is likely to be different in canted-AFM as compared to a routine FM/AFM.  The process of field cooling from above the $T{_N}$ leads to imprinting of canted-AFM domains guided by the relative strength of various energy scales including  exchange, Zeeman, magnetocrystalline anisotropy and DMI \cite{Binek, AK1}. Below a critical magnetic field, the canted-domains are formed by interplay between  intrinsic DMI and Zeeman. Above a critical H, the Zeeman term dominates over DMI and the AFM domains are guided by the applied magnetic field  rather than the DMI driven spin-canting.  Thus, at very high magnetic field, the quasi static remanence vanishes.  This seems to be related to unusual magnetic field dependence, which is reflected clearly in remanence, rather than \textit{in-field} magnetization.  These data indicate  that at low fields the phenomenon of  spin-canting  is  exclusively related to  the \textit{quasi static} part of the  remanence. 
	
\begin{figure}[!t]
\centering
\includegraphics[width=0.4\textwidth]{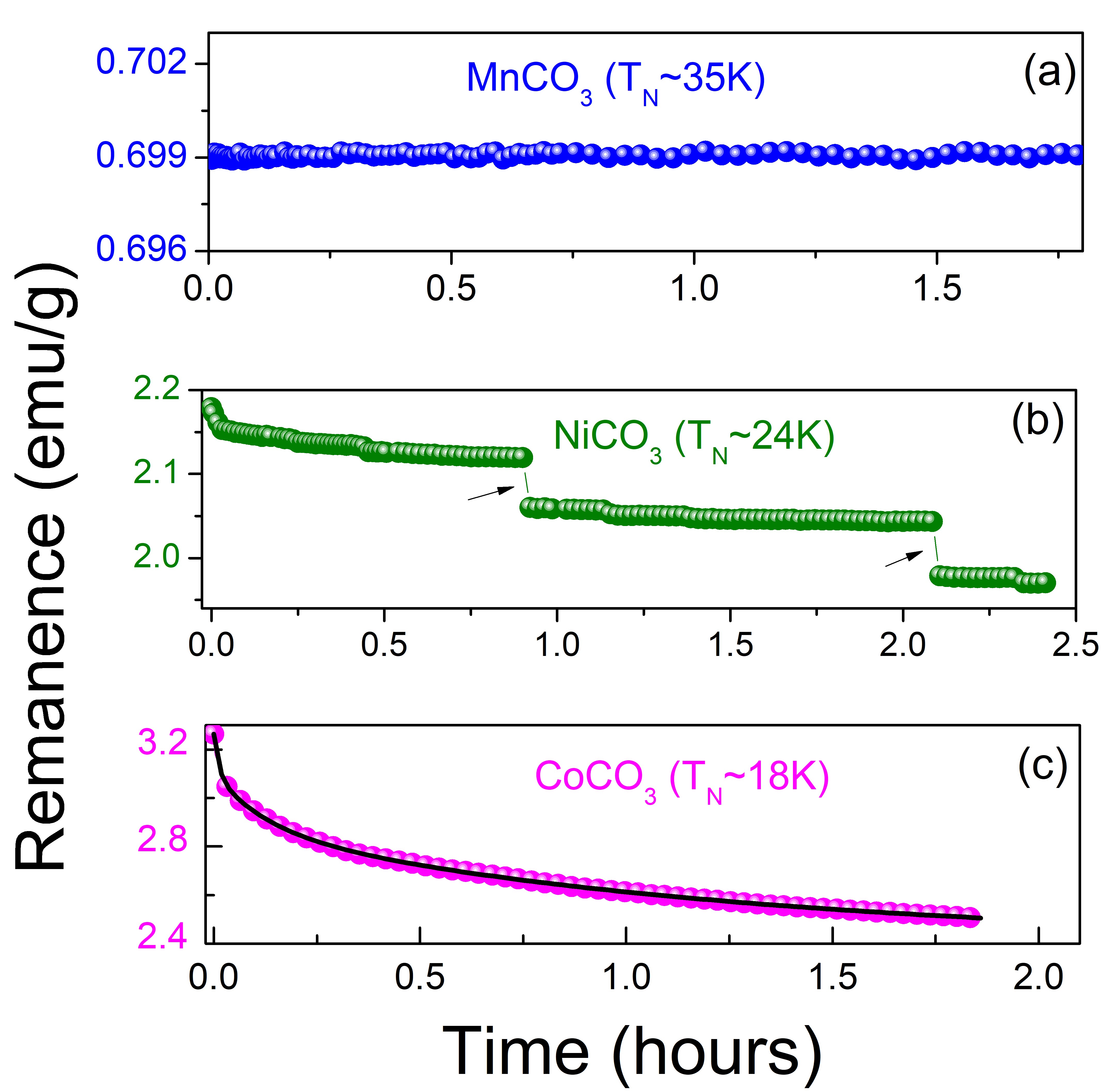}
\caption {\textbf{(a)}-\textbf{(c)} compares time-dependence of remanence at a fix temperature of 5K for  all three carbonates. The remanent state in all three cases has been prepared at a fixed H =100 Oe. The stability of \textit{quasi-static} remanence is seen to be highest in MnCO$_3$, which has highest T$_{N}$ in this series. The magnitude of  the remanence is also lowest in this case. The remanence in case of NiCO$_3$ exhibits avalanches like features, marked by arrows in \textbf{(b)}. These features are exclusive to  NiCO$_3$ and  reproducible when the remanent state is prepared in different magnetic fields.  CoCO$_{3}$, with lowest T$_{N}$ the $\mu$ vs \textit{time} follows the triple exponential decay, as shown with the black solid line in \textbf{(c)}.}
\label{Figure4}
\end{figure}

	It is to be noted that CoCO$_3$ also shows a tendency to peak like behavior, but it does not eventually vanish with increasing H, unlike the other two carbonates (Figure 3a). Here we recall that the spin-canting in CoCO$_{3}$ is known to have contributions from  DMI as well as SIA.\cite{Moriya1, Morrish, Radhakrishna, Martynov, Bazhan, Romanov1}.  The orbital magnetic moment in CoCO$_{3}$ remains relatively more unquenched \cite{Pincini, Loktev}  as compared to MnCO$_{3}$ and NiCO$_{3}$.  While DMI driven-spin canting relates to  \textit{T}$_N$ through J, the exchange coupling constant, the SIA is understood to be independent of J and hence \textit{T}$_N$  \cite{Moriya1}. With SIA  also contributing to spin-canting  (as is case of CoCO$_{3}$) it appears that the process of  field-cooling from above the  \textit{T}$_N$ may lead the canted-domains to still remained  pinned, when the remanant state is prepared in higher H.  Whether $\mu$ not vanishing with higher magnetic fields  in CoCO$_3$ indeed relates  to SIA term being significant in CoCO$_3$ needs further investigations. This can be confirmed by  tracking the nature of remanence in  pure SIA driven canted-AFMs. 
	
	We  note that  the  magnitude of this  $\mu$ is systematically largest in the sample with lowest \textit{T}$_N$, which is CoCO$_3$  in this series. Spin-canting angle is larger in samples with lower T$_{N}$ \cite{Dzy1, Moriya1}  which in-turn relates  to the lower exchange coupling constant associated with the primary AFM phase.  Thus DMI driven spin-canting angle is expected to be larger in samples with lower \textit{T}$_N$.  It is evident that this feature is clearly reflected in \textit{quasi-static} remanence, rather than magnetization in this series as the magnitude of the \textit{quasi-static} remanence is largest in the sample with lowest \textit{T}$_N$.   Overall, these data  establish that  all three samples exhibit a  \textit{quasi-static} $\mu$  which is exclusively related to spin-canting phenomenon. 
	
\subsection{\label{sec:level}  Remanence vs \textit{time} in CoCO$_3$, NiCO$_3$ and MnCO$_3$  }

	The magnetization  relaxation (i.e the  time dependence of remanent magnetization)   is an important tool for magnetic characterization. The remanent state in such measurements is also set by  cooling (or heating) the sample through the transition temperature in various H, switching off the H  and tracking the \textit{time-dependence} of remanence. These measurements reveal  the intricacies of magnetization dynamics in routine as well as  complex magnetic systems \cite{Kleemann, Benitez1, Benitez2}.  We also  measured magnetization relaxation phenomenon by tracking the time-dependence of $\mu$ prepared in different (cooling) fields. We first prepare a particular remanent state by following the similar experimental protocol outlined in Figure~\ref{Figure2}(a). After the $M{_{FC}}$ vs T run, the magnetic field is switched off at 5K and  the variations of $\mu$ with time (\textit{t}) is measured. The $\mu$ vs \textit{t} data  clearly bring forward the \textit{quasi-static} nature of remanence. For each sample, time dependence of $\mu$ was recorded at several (cooling) magnetic fields. For the sake of comparison, the data obtained for the identical (cooling) field of 100 Oe has been plotted for all three carbonates.  The magnetization relaxation in the time span of 1.5 to  2 hours for all three samples is compared in Figure~\ref{Figure4}. As is evident from Figure~\ref{Figure4}(a) the remanence exhibits hardly any decay  with time in case of  MnCO$_{3}$, the sample with \textit{T}$_N$ about 35K.   Similar features are also seen in case of  NiCO$_{3}$.  However, the  magnetization relaxation in the case of NiCO$_{3}$ proceeds through several discrete jumps, each of which is \textit{quasi-static} in nature, Figure~\ref{Figure4}(b).  Several measurements were repeated in  different (cooling)  magnetic fields which confirm that these  jumps are intrinsic to  NiCO$_{3}$. We also emphasize that such jumps in \textit{quasi-static} remanence were also observed in single-crystal hematite \cite{Pattanayak} as well as when hematite is encapsulated inside carbon nanotubes \cite{AK1}. The observation of such jumps appears to indicate the depinning of domains which are exclusively related to canted phase.  Thus, the existence of \textit{quasi-static} remanence exclusively connects to the DMI driven canted AFM for both MnCO$_{3}$ and NiCO$_{3}$. The stability of this \textit{quasi-static} remanence relates to the Neel temperature and the critical magnetic field that is applied during the preparation of the remanent state.   CoCO$_{3}$  on the other hand exhibits a mixed contribution from \textit{quasi-static} $\mu$  along with some relatively faster time scales, Figure4(c).  This is concluded from fitting the relaxation data to a triple exponential decay function, as shown in Figure~\ref{Figure4}(c). 

\begin{equation} \mu (t) = \mu_1 e^{-\frac{t}{\tau_1}} + \mu_2 e^{-\frac{t}{\tau_2}} + \mu_3 e^{-\frac{t}{\tau_3}}, \end{equation} 

 Here  $\tau_{1}$, $\tau_{2}$ and $\tau_{3}$ denote the three decay constants respectively. The solid line  in the main panel represent the triple exponential fits to the $\mu_{FC}$ versus time data. The decay times extracted for the particular \textit{H} of 100 Oe yield $\tau_{1}$ (= 0.01 h) $\textless$ $\tau_{2}$ (= 0.12 h) $\textless$ $\tau_{3}$ (= 1.22 h).  This implies  that  one of the time scale is ultra slow, with $\tau$  in the scale of hours.  While CoCO$_{3}$ also exhibits ultra -slow magnetization relaxation that leads to the observation of$\mu$, the additional features as observed in this case  appears to be related to the contribution of SIA and its strength as compared to other energy scales, including the exchange and Zeeman.  Again, relaxation data on pure SIA driven systems will  be illuminating in this context .

  Overall, we conclude that the stability of remanence with time reduces systematically for MnCO$_{3}$  NiCO$_{3}$ and  CoCO$_{3}$, that possess progressively decreasing \textit{T}$_N$.   On the other hand, the magnitude of the \textit{quasi-static} remanence  remains highest for sample with lowest \textit{T}$_N$, which is CoCO$_{3}$, but with a caveat that this sample also has contribution of SIA. Our core  observation  pertaining to magnitude of $\mu$ is  consistent with the fact that the  spin-canting angle increases  with decreasing  J and hence it is larger for samples with lower \textit{T}$_N$, especially for DMI driven systems.   For  SIA driven systems, spin-canting angle  should not vary significantly with  J and therefore with  \textit{T}$_N$. While primary mechanism of spin canting in CoCO$_{3}$ is DMI, it is known to have contribution from SIA as well. Experimental and theoretical data has revealed that the spin canting angle is  larger in NiCO$_{3}$ as compared to CoCO$_{3}$ \cite{Beutier}.   The manifestation of SIA related factors  appear to be reflected in the \textit{time} as well as \textit{magnetic-field} dependence of  $\mu$ of CoCO$_{3}$.   Whether this dynamics captures the essential physics behind the canting mechanism (DMI, SIA or both) needs a comparative  study on pure SIA-driven canted-AFM. From the present set of data, we expect \textit{quasi-static} remanence to exist in spin-canted systems,  but the \textit{field} and  \textit{time} dependence of this \textit{quasi-static} remanence  may  be different in DMI driven systems from the SIA driven ones. Exploring the functional form of $\mu$  can thus provide insights  in isolating the DMI or SIA contribution in new systems. 
	
		\begin{figure}[!t]
\centering
\includegraphics[width=0.5\textwidth]{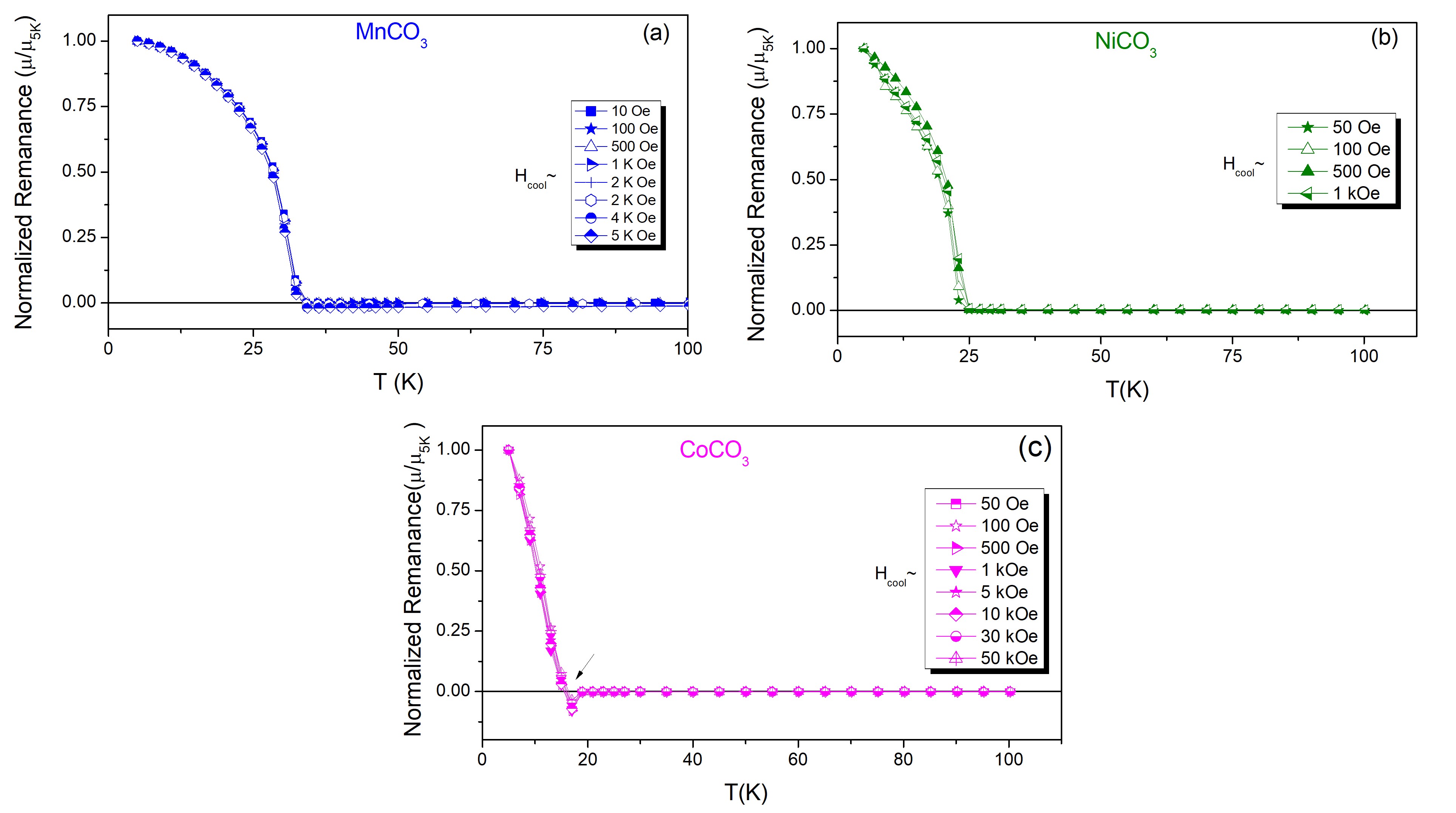}
\caption { \textbf{(a)} - \textbf{(c)} show normalized $\mu$  vs T  exhibiting scaling in remanence at selected  \textit{H}  for all three carbonates and relate to the intrinsic phenomenon of piezomagnetism.}
\label{Figure5}
\end{figure} 

\subsection{\label{sec:level} Scaling  of \textit{quasi-static} Remanence in CoCO$_3$, NiCO$_3$ and MnCO$_3$ }

	While the counter-intuitive magnetic field dependence along with \textit{quasi-static} nature of remanence is observed in all three carbonates, which are symmetry allowed spin-canted systems, there is an associated phenomenon of piezomagnetism, which is equally intriguing\cite{Romanov1, Dzy2, Landau, Attfield}.  There have been reports from other groups that reveal a scaling phenomenon in remanence measurements in systems in which piezomagnetism arises  from size-effects or doping \cite{Binek,Kleemann}.  This scaling is understood to arise from  the observation that the remanence factorizes  with temperature and magnetic field variables following the equation  $\mu_{FC}$ (\textit{H},\textit{T}) = f(\textit{H}) g(\textit{T}))\cite{Binek,Kleemann}. This factorization  leads to a universal scaling, when  normalized remanences (prepared  in different (cooling)  magnetic fields) are plotted as a function of T.  We have  earlier observed this scaling phenomenon  in a chrome -oxide based hybrid system in which the individual constituent is not a symmetry allowed piezomagnet, but arises due to interface effects \cite{Ashna1}.  In present case, all three carbonates are also symmetry allowed piezomagnets. In light of the fact that the remanence in CoCO$_{3}$  is almost self-scaled, we were tempted to explore scaling effects in remaining two carbonates.  These data are  shown in Figure~\ref{Figure5}(a)-(c) for all three samples.  Here $\mu$ vs T runs,  in which the magnitude of $\mu$  that is not vanishingly small, are chosen. The rationale is that  when $\mu$  is negligibly  small, the data are  influenced by the residual magnetic field associated with the superconducting magnet of SQUID. This residual field is often negative and can lead to a residual (and negative) magnetization, as is highlighted with a black arrow in Figure~\ref{Figure5}(c) and also in Figure~\ref{Figure2}(d). This negative residual magnetization  also becomes prominent in the vicinity of \textit{T}$_N$, when $\mu$ is about to vanish, as is evident from Figure~\ref{Figure2}(d) and Figure~\ref{Figure5}(c). Barring such runs, when we plot normalized $\mu$ vs T at different (cooling) H,  we do see a  universal scaling in \textit{quasi-static} remanence  for all three samples. This observation is also consistent with previous reports on the scaling of \textit{quasi-static}  remanence  in which the effect arises due to size and interface effects\cite{Ashna1, Ashna2}. This universal scaling as observed in all three carbonates confirms the concurrent phenomenon of piezomagnetism in spin-canted AFM .  
	
	\subsection{\label{sec:level} Neutron Diffraction in remanent state in CoCO$_{3}$ }

\begin{figure}[!t]
\centering
\includegraphics[width=0.5\textwidth]{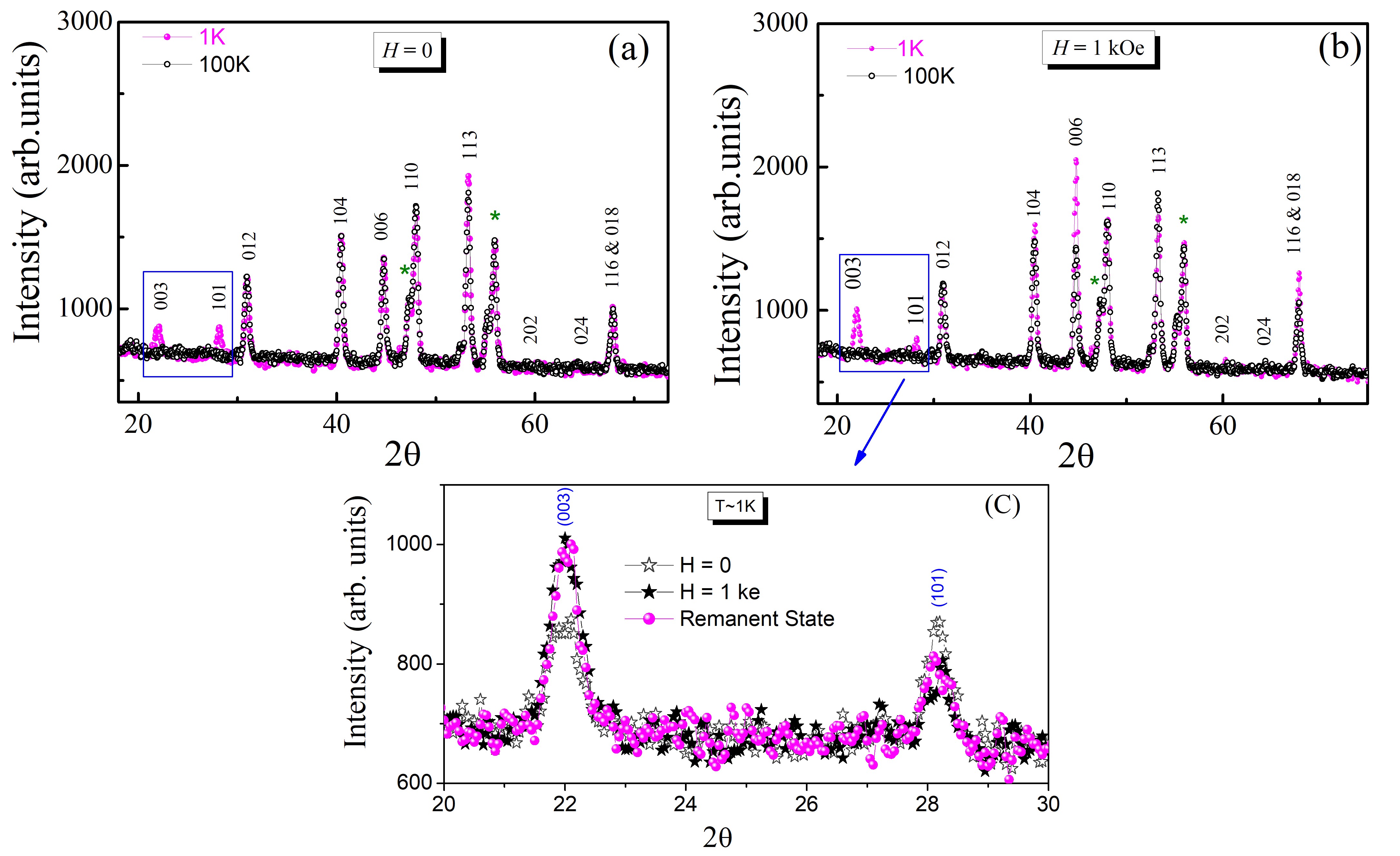}
\caption { \textbf{(a)} and \textbf{(b)} compares  neutron powder diffraction data (NPD)  at 100 K and 1K  measured in zero magnetic field and in 1kOe  respectively. The black  dots in both cases represent NPD in paramagnetic region (100K) and pink dots show the same  antiferromagnetic region (1K).  The indices are written with respect to the R-3c space group in hex setting. The pure AFM peaks (003) \& (110) are highlighted in \textbf{(a)} and \textbf{(b)}. Data in \textbf{(c)} highlights NPD at 1K in zero H (open black stars) , 1kOe (filled black stars) and in corresponding remanent state (pink dots) for CoCO$_{3}$ for pure AFM peaks.}
\label{Figure6}
\end{figure}

	 To gain microscopic evidence for the ultra-slow relaxation phenomena that leads to \textit{quasi-static} remanence  in these DMI driven systems, we performed neutron powder diffraction (NPD) measurements on a representative sample, CoCO$_{3}$. We have recorded the NPD patterns at two selected temperatures, 100 K and 1K. Considering that the \textit{T}$_N$ of CoCO$_{3}$ is 18 K, the NPD recorded at 100 K covers the paramagnetic region and that recoded at 1K is in the AFM region. The data is recorded at these two fixed temperatures in (i) zero H (ii) in presence of H = 2kOe (iii) in the corresponding remanent state. These data are presented in Figure~\ref{Figure6}.
	
	The sample is first cooled in zero magnetic field from the room temperature and NPD  is recorded at 100 K (in the paramagnetic region for CoCO$_{3}$). This  is shown as open black circles in Figure~\ref{Figure6}(a). The peak indexing in Fig. 6(a) is consistent with the Plumier's work on neutron diffraction study on the isostructural compound NiCO$_{3}$ \cite{Plumier1, Plumier2}. The indices in the NPD pattern are written with respect to R-3c space group in the hexagonal setting. The peaks marked with an asterisk  arise from the cryomagnet during NPD measurements. It is worth mentioning here that there are no impurity peaks in XRD pattern of the CoCO$_{3}$ sample (see supplemental information, Text S3 for details).
	
	After recording NPD at 100 K, the sample was further cooled to  even lower temperatures  in zero field and the NPD was recorded at 1K (below the T$_{N}$ of CoCO$_{3}$ ). These data are  shown as  pink- dots in Figure~\ref{Figure6}(a). It is evident from the comparison of the NPD profiles recorded at 1K and 100 K  in Figure~\ref{Figure6}(a),  the two additional peaks (003 and 101) other than the main perovskite peaks are present in the 1K NPD patterns.  These purely magnetic  peaks (003) and (101)  at 1K are highlighted in Figure~\ref{Figure6}(a), as they are exclusive to  AFM phase of CoCO$_{3}$.  The additional magnetic Bragg peaks at 1K  are indexed with a propagation vector k = (0,0,0) corresponding to the space group R-3c. The magnetic structures compatible with the R-3c symmetry are determined by representation analysis technique. For the propagation vector k = (0,0,0), the magnetic reducible representation  can be decomposed as direct sum of three irreducible representations, out of which only one explains the observed magnetic peaks and spin-canting model. For the sake of clarity, this detailed Rietveld profile refinement of the NPD data at each temperature and magnetic field is presented  in supp. information Figure S3(a)-(d). The magnetic structure is also schematically shown in supp. Info, Figure S4. 
	
	 For recording the \textit{in-field} and the corresponding \textit{remanent-state NPD}, the sample was again cooled from the room temperature down to 100 K in the presence of 1 kOe magnetic field.  The  NPD was recorded at T= 100K in the paramagnetic region of CoCO$_{3}$ , in presence of 1kOe.  These data  are shown as black circles in Figure~\ref{Figure6}(b).  The sample was then cooled down to to 1K and the NPD was recorded in presence of 1 kOe magnetic field. This patterns is shown as pink dots in Figure~\ref{Figure6}(b). We note from the Figure~\ref{Figure6}(b) that in the presence of magnetic field, the intensity of the strongest magnetic peak (003) increases significantly, as compared to the zero-field NPD pattern. Subsequently, the magnetic field is switched-off and  NPD is  now  recorded in remanent state, while temperature is still held at 1K. It is to be noted that the entire protocol of measuring this remanent state NPD is similar to what is adopted  to track remanence using SQUID magnetometry.

	We finally compare NPD pattern at 1K in zero magnetic field, in presence of 1kOe magnetic field and in the corresponding remanent state in Figure~\ref{Figure6}(c). Here, for the sake of clarity  we focus on (003) and (101) peaks, which are  pure AFM peaks. Figure~\ref{Figure6}(c)  clearly shows that the intensity of the strongest peak (003) in the presence of  magnetic field  and in the remanent state are comparable. This suggests that the NPD data not only show the relative change in intensity in the presence of H, but also confirms that the intensity is retained in the remanent state. Microscopically, these data demonstrate  that once the canted domains have been guided during the field cooling process, their rearrangement is energetically unfavorable after the  magnetic field is removed. This feature in NPD relates to the \textit{quasi-static} remanence, which we observe in bulk magnetization through SQUID magnetometry.  The NPD data further  confirm the ultra-slow magnetization dynamics associated with canted AFM, which is also reflected in the bulk magnetometry.  Thus, in addition to our bulk magnetization measurements, NPD data further supports the presence of  \textit{quasi-static} remanence and its connection with spin-canting.  
	
\section{Conclusions}

	In conclusion, we have conducted remanence measurements in a series of carbonates which are prototypical canted-antiferromagnets and piezomagnets with systematically varying  \textit{T}$_N$. These are primarily Dzyaloshiskii Moriya Interaction driven systems. However,  contribution of single in anisotropy is significant for  CoCO$_{3}$, as compared to other two carbonates.  All three samples exhibit a unique feature in remanence, a part which is \textit{quasi-static} in nature and  exhibits a counter-intuitive magnetic field dependence. The magnitude  of this  \textit{quasi-static} remanence  shows a systematic variation with  \textit{T}$_N$ and  therefore with the extent of spin-canting.  The magnitude is  larger for sample with  smaller \textit{T}$_N$,  which is  CoCO$_{3}$ in this series of compounds. The stability of remanence with respect to time is higher for samples with higher  \textit{T}$_N$, which is  MnCO$_{3}$.  NiCO$_{3}$  shows a similar feature, albeit with discrete jumps.  The normalized remanence also exhibits  a  universal scaling  for all three samples,  which appears to be associated with  the concurrent phenomenon of piezomagnetism. The existence of this \textit{quasi-static} remanence is also seen in neutron diffraction, conducted in remanent state. The specifics related to the variation of  \textit{quasi -static} remanence  with \textit{magnetic field} or \textit{time} can provide crucial information  for  isolating the  Dzyaloshinskii Moriya Interaction driven canted-antiferromagnets  from single-ion-anisotropy driven ones.

\section*{Supplementary Material}
See the Supplementary Information for synthesis details of Mn, Ni and CoCO$_{3}$ in Text S1 \& Text S2.  X-Ray diffraction patterns along with the Rietveld Profile refinement of XRD data  and low field MH  for all three carbonates is given as Figure S1 and Figure S2 respectively. Neutron diffraction data along with  Rietveld profile refinements and magnetic structure details  of sample CoCO$_{3}$ in section III of SI. 

\section*{Acknowledgements}	

	SN thanks the Department of Science and Technology, India (SR/NM/Z-07/2015) for financial support to carry out the neutron diffraction experiment, and Jawaharlal Nehru Centre for Advanced Scientific Research (JNCASR) for managing the project. AB acknowledges DST-SERB grant, India (CRG/2022/008373) for funding support.
	
\nocite{*}
\bibliography{Ref}

\begin{thebibliography}{71}%
\makeatletter
\providecommand \@ifxundefined [1]{%
 \@ifx{#1\undefined}
}%
\providecommand \@ifnum [1]{%
 \ifnum #1\expandafter \@firstoftwo
 \else \expandafter \@secondoftwo
 \fi
}%
\providecommand \@ifx [1]{%
 \ifx #1\expandafter \@firstoftwo
 \else \expandafter \@secondoftwo
 \fi
}%
\providecommand \natexlab [1]{#1}%
\providecommand \enquote  [1]{``#1''}%
\providecommand \bibnamefont  [1]{#1}%
\providecommand \bibfnamefont [1]{#1}%
\providecommand \citenamefont [1]{#1}%
\providecommand \href@noop [0]{\@secondoftwo}%
\providecommand \href [0]{\begingroup \@sanitize@url \@href}%
\providecommand \@href[1]{\@@startlink{#1}\@@href}%
\providecommand \@@href[1]{\endgroup#1\@@endlink}%
\providecommand \@sanitize@url [0]{\catcode `\\12\catcode `\$12\catcode
  `\&12\catcode `\#12\catcode `\^12\catcode `\_12\catcode `\%12\relax}%
\providecommand \@@startlink[1]{}%
\providecommand \@@endlink[0]{}%
\providecommand \url  [0]{\begingroup\@sanitize@url \@url }%
\providecommand \@url [1]{\endgroup\@href {#1}{\urlprefix }}%
\providecommand \urlprefix  [0]{URL }%
\providecommand \Eprint [0]{\href }%
\providecommand \doibase [0]{https://doi.org/}%
\providecommand \selectlanguage [0]{\@gobble}%
\providecommand \bibinfo  [0]{\@secondoftwo}%
\providecommand \bibfield  [0]{\@secondoftwo}%
\providecommand \translation [1]{[#1]}%
\providecommand \BibitemOpen [0]{}%
\providecommand \bibitemStop [0]{}%
\providecommand \bibitemNoStop [0]{.\EOS\space}%
\providecommand \EOS [0]{\spacefactor3000\relax}%
\providecommand \BibitemShut  [1]{\csname bibitem#1\endcsname}%
\let\auto@bib@innerbib\@empty
\bibitem [{\citenamefont {Binz}\ \emph {et~al.}(2006)\citenamefont {Binz},
  \citenamefont {Vishwanath},\ and\ \citenamefont {Aji}}]{Binz}%
  \BibitemOpen
  \bibfield  {author} {\bibinfo {author} {\bibfnamefont {B.}~\bibnamefont
  {Binz}}, \bibinfo {author} {\bibfnamefont {A.}~\bibnamefont {Vishwanath}},\
  and\ \bibinfo {author} {\bibfnamefont {V.}~\bibnamefont {Aji}},\ }\bibfield
  {title} {\bibinfo {title} {Theory of the helical spin crystal: A candidate
  for the partially ordered state of mnsi},\ }\href
  {https://doi.org/10.1103/PhysRevLett.96.207202} {\bibfield  {journal}
  {\bibinfo  {journal} {Phys. Rev. Lett.}\ }\textbf {\bibinfo {volume} {96}},\
  \bibinfo {pages} {207202} (\bibinfo {year} {2006})}\BibitemShut {NoStop}%
\bibitem [{\citenamefont {Gayles}\ \emph {et~al.}(2015)\citenamefont {Gayles},
  \citenamefont {Freimuth}, \citenamefont {Schena}, \citenamefont {Lani},
  \citenamefont {Mavropoulos}, \citenamefont {Duine}, \citenamefont {Bl\"ugel},
  \citenamefont {Sinova},\ and\ \citenamefont {Mokrousov}}]{Gayles}%
  \BibitemOpen
  \bibfield  {author} {\bibinfo {author} {\bibfnamefont {J.}~\bibnamefont
  {Gayles}}, \bibinfo {author} {\bibfnamefont {F.}~\bibnamefont {Freimuth}},
  \bibinfo {author} {\bibfnamefont {T.}~\bibnamefont {Schena}}, \bibinfo
  {author} {\bibfnamefont {G.}~\bibnamefont {Lani}}, \bibinfo {author}
  {\bibfnamefont {P.}~\bibnamefont {Mavropoulos}}, \bibinfo {author}
  {\bibfnamefont {R.~A.}\ \bibnamefont {Duine}}, \bibinfo {author}
  {\bibfnamefont {S.}~\bibnamefont {Bl\"ugel}}, \bibinfo {author}
  {\bibfnamefont {J.}~\bibnamefont {Sinova}},\ and\ \bibinfo {author}
  {\bibfnamefont {Y.}~\bibnamefont {Mokrousov}},\ }\bibfield  {title} {\bibinfo
  {title} {Dzyaloshinskii-moriya interaction and hall effects in the skyrmion
  phase of ${\mathrm{mn}}_{1\ensuremath{-}x}{\mathrm{fe}}_{x}\mathrm{Ge}$},\
  }\href {https://doi.org/10.1103/PhysRevLett.115.036602} {\bibfield  {journal}
  {\bibinfo  {journal} {Phys. Rev. Lett.}\ }\textbf {\bibinfo {volume} {115}},\
  \bibinfo {pages} {036602} (\bibinfo {year} {2015})}\BibitemShut {NoStop}%
\bibitem [{\citenamefont {Sinova}\ and\ \citenamefont {Zutic}(2012)}]{Jairo}%
  \BibitemOpen
  \bibfield  {author} {\bibinfo {author} {\bibfnamefont {J.}~\bibnamefont
  {Sinova}}\ and\ \bibinfo {author} {\bibfnamefont {I.}~\bibnamefont {Zutic}},\
  }\href@noop {} {\bibfield  {journal} {\bibinfo  {journal} {Nature Materials}\
  }\textbf {\bibinfo {volume} {11}},\ \bibinfo {pages} {368} (\bibinfo {year}
  {2012})}\BibitemShut {NoStop}%
\bibitem [{\citenamefont {Baltz}\ \emph {et~al.}(2018)\citenamefont {Baltz},
  \citenamefont {Manchon}, \citenamefont {Tsoi}, \citenamefont {Moriyama},
  \citenamefont {Ono},\ and\ \citenamefont {Tserkovnyak}}]{Baltz}%
  \BibitemOpen
  \bibfield  {author} {\bibinfo {author} {\bibfnamefont {V.}~\bibnamefont
  {Baltz}}, \bibinfo {author} {\bibfnamefont {A.}~\bibnamefont {Manchon}},
  \bibinfo {author} {\bibfnamefont {M.}~\bibnamefont {Tsoi}}, \bibinfo {author}
  {\bibfnamefont {T.}~\bibnamefont {Moriyama}}, \bibinfo {author}
  {\bibfnamefont {T.}~\bibnamefont {Ono}},\ and\ \bibinfo {author}
  {\bibfnamefont {Y.}~\bibnamefont {Tserkovnyak}},\ }\bibfield  {title}
  {\bibinfo {title} {Antiferromagnetic spintronics},\ }\href
  {https://doi.org/10.1103/RevModPhys.90.015005} {\bibfield  {journal}
  {\bibinfo  {journal} {Rev. Mod. Phys.}\ }\textbf {\bibinfo {volume} {90}},\
  \bibinfo {pages} {015005} (\bibinfo {year} {2018})}\BibitemShut {NoStop}%
\bibitem [{\citenamefont {Ajejas}\ \emph {et~al.}(2018)\citenamefont {Ajejas},
  \citenamefont {Gud{\'\i}n}, \citenamefont {Guerrero}, \citenamefont
  {Anad{\'o}n~Barcelona}, \citenamefont {Diez}, \citenamefont {de~Melo~Costa},
  \citenamefont {Olleros}, \citenamefont {Ni{\~n}o}, \citenamefont {Pizzini},
  \citenamefont {Vogel} \emph {et~al.}}]{Ajejas}%
  \BibitemOpen
  \bibfield  {author} {\bibinfo {author} {\bibfnamefont {F.}~\bibnamefont
  {Ajejas}}, \bibinfo {author} {\bibfnamefont {A.}~\bibnamefont {Gud{\'\i}n}},
  \bibinfo {author} {\bibfnamefont {R.}~\bibnamefont {Guerrero}}, \bibinfo
  {author} {\bibfnamefont {A.}~\bibnamefont {Anad{\'o}n~Barcelona}}, \bibinfo
  {author} {\bibfnamefont {J.~M.}\ \bibnamefont {Diez}}, \bibinfo {author}
  {\bibfnamefont {L.}~\bibnamefont {de~Melo~Costa}}, \bibinfo {author}
  {\bibfnamefont {P.}~\bibnamefont {Olleros}}, \bibinfo {author} {\bibfnamefont
  {M.~A.}\ \bibnamefont {Ni{\~n}o}}, \bibinfo {author} {\bibfnamefont
  {S.}~\bibnamefont {Pizzini}}, \bibinfo {author} {\bibfnamefont
  {J.}~\bibnamefont {Vogel}}, \emph {et~al.},\ }\bibfield  {title} {\bibinfo
  {title} {Unraveling dzyaloshinskii--moriya interaction and chiral nature of
  graphene/cobalt interface},\ }\href@noop {} {\bibfield  {journal} {\bibinfo
  {journal} {Nano letters}\ }\textbf {\bibinfo {volume} {18}},\ \bibinfo
  {pages} {5364} (\bibinfo {year} {2018})}\BibitemShut {NoStop}%
\bibitem [{\citenamefont {Di}\ \emph {et~al.}(2015)\citenamefont {Di},
  \citenamefont {Zhang}, \citenamefont {Lim}, \citenamefont {Ng}, \citenamefont
  {Kuok}, \citenamefont {Yu}, \citenamefont {Yoon}, \citenamefont {Qiu},\ and\
  \citenamefont {Yang}}]{Di}%
  \BibitemOpen
  \bibfield  {author} {\bibinfo {author} {\bibfnamefont {K.}~\bibnamefont
  {Di}}, \bibinfo {author} {\bibfnamefont {V.~L.}\ \bibnamefont {Zhang}},
  \bibinfo {author} {\bibfnamefont {H.~S.}\ \bibnamefont {Lim}}, \bibinfo
  {author} {\bibfnamefont {S.~C.}\ \bibnamefont {Ng}}, \bibinfo {author}
  {\bibfnamefont {M.~H.}\ \bibnamefont {Kuok}}, \bibinfo {author}
  {\bibfnamefont {J.}~\bibnamefont {Yu}}, \bibinfo {author} {\bibfnamefont
  {J.}~\bibnamefont {Yoon}}, \bibinfo {author} {\bibfnamefont {X.}~\bibnamefont
  {Qiu}},\ and\ \bibinfo {author} {\bibfnamefont {H.}~\bibnamefont {Yang}},\
  }\bibfield  {title} {\bibinfo {title} {Direct observation of the
  dzyaloshinskii-moriya interaction in a pt/co/ni film},\ }\href
  {https://doi.org/10.1103/PhysRevLett.114.047201} {\bibfield  {journal}
  {\bibinfo  {journal} {Phys. Rev. Lett.}\ }\textbf {\bibinfo {volume} {114}},\
  \bibinfo {pages} {047201} (\bibinfo {year} {2015})}\BibitemShut {NoStop}%
\bibitem [{\citenamefont {N{\v{e}}mec}\ \emph {et~al.}(2018)\citenamefont
  {N{\v{e}}mec}, \citenamefont {Fiebig}, \citenamefont {Kampfrath},\ and\
  \citenamefont {Kimel}}]{Nvemec}%
  \BibitemOpen
  \bibfield  {author} {\bibinfo {author} {\bibfnamefont {P.}~\bibnamefont
  {N{\v{e}}mec}}, \bibinfo {author} {\bibfnamefont {M.}~\bibnamefont {Fiebig}},
  \bibinfo {author} {\bibfnamefont {T.}~\bibnamefont {Kampfrath}},\ and\
  \bibinfo {author} {\bibfnamefont {A.~V.}\ \bibnamefont {Kimel}},\ }\bibfield
  {title} {\bibinfo {title} {Antiferromagnetic opto-spintronics},\ }\href@noop
  {} {\bibfield  {journal} {\bibinfo  {journal} {Nature Physics}\ }\textbf
  {\bibinfo {volume} {14}},\ \bibinfo {pages} {229} (\bibinfo {year}
  {2018})}\BibitemShut {NoStop}%
\bibitem [{\citenamefont {Weingart}\ \emph {et~al.}(2012)\citenamefont
  {Weingart}, \citenamefont {Spaldin},\ and\ \citenamefont
  {Bousquet}}]{Nicola}%
  \BibitemOpen
  \bibfield  {author} {\bibinfo {author} {\bibfnamefont {C.}~\bibnamefont
  {Weingart}}, \bibinfo {author} {\bibfnamefont {N.}~\bibnamefont {Spaldin}},\
  and\ \bibinfo {author} {\bibfnamefont {E.}~\bibnamefont {Bousquet}},\
  }\bibfield  {title} {\bibinfo {title} {Noncollinear magnetism and single-ion
  anisotropy in multiferroic perovskites},\ }\href
  {https://doi.org/10.1103/PhysRevB.86.094413} {\bibfield  {journal} {\bibinfo
  {journal} {Phys. Rev. B}\ }\textbf {\bibinfo {volume} {86}},\ \bibinfo
  {pages} {094413} (\bibinfo {year} {2012})}\BibitemShut {NoStop}%
\bibitem [{\citenamefont {Dzyaloshinskii}(1957{\natexlab{a}})}]{Dzy1}%
  \BibitemOpen
  \bibfield  {author} {\bibinfo {author} {\bibfnamefont {I.~E.}\ \bibnamefont
  {Dzyaloshinskii}},\ }\href@noop {} {\bibfield  {journal} {\bibinfo  {journal}
  {JETP}\ }\textbf {\bibinfo {volume} {32}},\ \bibinfo {pages} {1259} (\bibinfo
  {year} {1957}{\natexlab{a}})}\BibitemShut {NoStop}%
\bibitem [{\citenamefont {Moriya}(1960)}]{Moriya1}%
  \BibitemOpen
  \bibfield  {author} {\bibinfo {author} {\bibfnamefont {T.}~\bibnamefont
  {Moriya}},\ }\href {https://doi.org/10.1103/PhysRev.120.91} {\bibfield
  {journal} {\bibinfo  {journal} {Phys. Rev.}\ }\textbf {\bibinfo {volume}
  {120}},\ \bibinfo {pages} {91} (\bibinfo {year} {1960})}\BibitemShut
  {NoStop}%
\bibitem [{\citenamefont {Pincini}\ \emph {et~al.}(2018)\citenamefont
  {Pincini}, \citenamefont {Fabrizi}, \citenamefont {Beutier}, \citenamefont
  {Nisbet}, \citenamefont {Elnaggar}, \citenamefont {Dmitrienko}, \citenamefont
  {Katsnelson}, \citenamefont {Kvashnin}, \citenamefont {Lichtenstein},
  \citenamefont {Mazurenko}, \citenamefont {Ovchinnikova}, \citenamefont
  {Dimitrova},\ and\ \citenamefont {Collins}}]{Pincini}%
  \BibitemOpen
  \bibfield  {author} {\bibinfo {author} {\bibfnamefont {D.}~\bibnamefont
  {Pincini}}, \bibinfo {author} {\bibfnamefont {F.}~\bibnamefont {Fabrizi}},
  \bibinfo {author} {\bibfnamefont {G.}~\bibnamefont {Beutier}}, \bibinfo
  {author} {\bibfnamefont {G.}~\bibnamefont {Nisbet}}, \bibinfo {author}
  {\bibfnamefont {H.}~\bibnamefont {Elnaggar}}, \bibinfo {author}
  {\bibfnamefont {V.~E.}\ \bibnamefont {Dmitrienko}}, \bibinfo {author}
  {\bibfnamefont {M.~I.}\ \bibnamefont {Katsnelson}}, \bibinfo {author}
  {\bibfnamefont {Y.~O.}\ \bibnamefont {Kvashnin}}, \bibinfo {author}
  {\bibfnamefont {A.~I.}\ \bibnamefont {Lichtenstein}}, \bibinfo {author}
  {\bibfnamefont {V.~V.}\ \bibnamefont {Mazurenko}}, \bibinfo {author}
  {\bibfnamefont {E.~N.}\ \bibnamefont {Ovchinnikova}}, \bibinfo {author}
  {\bibfnamefont {O.~V.}\ \bibnamefont {Dimitrova}},\ and\ \bibinfo {author}
  {\bibfnamefont {S.~P.}\ \bibnamefont {Collins}},\ }\bibfield  {title}
  {\bibinfo {title} {Role of the orbital moment in a series of isostructural
  weak ferromagnets},\ }\href {https://doi.org/10.1103/PhysRevB.98.104424}
  {\bibfield  {journal} {\bibinfo  {journal} {Phys. Rev. B}\ }\textbf {\bibinfo
  {volume} {98}},\ \bibinfo {pages} {104424} (\bibinfo {year}
  {2018})}\BibitemShut {NoStop}%
\bibitem [{\citenamefont {Pattanayak}\ \emph {et~al.}(2017)\citenamefont
  {Pattanayak}, \citenamefont {Bhattacharyya}, \citenamefont {Nigam},
  \citenamefont {Cheong},\ and\ \citenamefont {Bajpai}}]{Pattanayak}%
  \BibitemOpen
  \bibfield  {author} {\bibinfo {author} {\bibfnamefont {N.}~\bibnamefont
  {Pattanayak}}, \bibinfo {author} {\bibfnamefont {A.}~\bibnamefont
  {Bhattacharyya}}, \bibinfo {author} {\bibfnamefont {A.~K.}\ \bibnamefont
  {Nigam}}, \bibinfo {author} {\bibfnamefont {S.-W.}\ \bibnamefont {Cheong}},\
  and\ \bibinfo {author} {\bibfnamefont {A.}~\bibnamefont {Bajpai}},\
  }\bibfield  {title} {\bibinfo {title} {Quasistatic remanence in
  dzyaloshinskii-moriya interaction driven weak ferromagnets and
  piezomagnets},\ }\href {https://doi.org/10.1103/PhysRevB.96.104422}
  {\bibfield  {journal} {\bibinfo  {journal} {Phys. Rev. B}\ }\textbf {\bibinfo
  {volume} {96}},\ \bibinfo {pages} {104422} (\bibinfo {year}
  {2017})}\BibitemShut {NoStop}%
\bibitem [{\citenamefont {Pattanayak}\ \emph {et~al.}(2019)\citenamefont
  {Pattanayak}, \citenamefont {Bhattacharyya}, \citenamefont {Chakravarty},\
  and\ \citenamefont {Bajpai}}]{Pattanayak1}%
  \BibitemOpen
  \bibfield  {author} {\bibinfo {author} {\bibfnamefont {N.}~\bibnamefont
  {Pattanayak}}, \bibinfo {author} {\bibfnamefont {A.}~\bibnamefont
  {Bhattacharyya}}, \bibinfo {author} {\bibfnamefont {S.}~\bibnamefont
  {Chakravarty}},\ and\ \bibinfo {author} {\bibfnamefont {A.}~\bibnamefont
  {Bajpai}},\ }\bibfield  {title} {\bibinfo {title} {Weak ferromagnetism and
  time-stable remanence in hematite: effect of shape, size and morphology},\
  }\href@noop {} {\bibfield  {journal} {\bibinfo  {journal} {Journal of
  Physics: Condensed Matter}\ }\textbf {\bibinfo {volume} {31}},\ \bibinfo
  {pages} {365802} (\bibinfo {year} {2019})}\BibitemShut {NoStop}%
\bibitem [{\citenamefont {Kapoor}\ \emph {et~al.}(2019)\citenamefont {Kapoor},
  \citenamefont {Dey}, \citenamefont {Garg},\ and\ \citenamefont
  {Bajpai}}]{AK1}%
  \BibitemOpen
  \bibfield  {author} {\bibinfo {author} {\bibfnamefont {A.}~\bibnamefont
  {Kapoor}}, \bibinfo {author} {\bibfnamefont {A.~B.}\ \bibnamefont {Dey}},
  \bibinfo {author} {\bibfnamefont {C.}~\bibnamefont {Garg}},\ and\ \bibinfo
  {author} {\bibfnamefont {A.}~\bibnamefont {Bajpai}},\ }\bibfield  {title}
  {\bibinfo {title} {Enhanced magnetism and time-stable remanence at the
  interface of hematite and carbon nanotubes},\ }\href
  {https://doi.org/10.1088/1361-6528/ab27ec} {\bibfield  {journal} {\bibinfo
  {journal} {Nanotechnology}\ }\textbf {\bibinfo {volume} {30}},\ \bibinfo
  {pages} {385706} (\bibinfo {year} {2019})}\BibitemShut {NoStop}%
\bibitem [{\citenamefont {Sahoo}\ and\ \citenamefont {Binek}(2007)}]{Binek}%
  \BibitemOpen
  \bibfield  {author} {\bibinfo {author} {\bibfnamefont {S.}~\bibnamefont
  {Sahoo}}\ and\ \bibinfo {author} {\bibfnamefont {C.}~\bibnamefont {Binek}},\
  }\bibfield  {title} {\bibinfo {title} {Piezomagnetism in epitaxial cr2o3 thin
  films and spintronic applications},\ }\href
  {https://doi.org/10.1080/09500830701253177} {\bibfield  {journal} {\bibinfo
  {journal} {Philosophical Magazine Letters}\ }\textbf {\bibinfo {volume}
  {87}},\ \bibinfo {pages} {259} (\bibinfo {year} {2007})}\BibitemShut
  {NoStop}%
\bibitem [{\citenamefont {Bajpai}\ \emph {et~al.}(2010)\citenamefont {Bajpai},
  \citenamefont {Klingeler}, \citenamefont {Wizent}, \citenamefont {Nigam},
  \citenamefont {Cheong},\ and\ \citenamefont {Büchner}}]{Ashna1}%
  \BibitemOpen
  \bibfield  {author} {\bibinfo {author} {\bibfnamefont {A.}~\bibnamefont
  {Bajpai}}, \bibinfo {author} {\bibfnamefont {R.}~\bibnamefont {Klingeler}},
  \bibinfo {author} {\bibfnamefont {N.}~\bibnamefont {Wizent}}, \bibinfo
  {author} {\bibfnamefont {A.~K.}\ \bibnamefont {Nigam}}, \bibinfo {author}
  {\bibfnamefont {S.-W.}\ \bibnamefont {Cheong}},\ and\ \bibinfo {author}
  {\bibfnamefont {B.}~\bibnamefont {Büchner}},\ }\bibfield  {title} {\bibinfo
  {title} {Unusual field dependence of remanent magnetization in granular
  $cro_{2}$ : the possible relevance of piezomagnetism},\ }\href
  {http://stacks.iop.org/0953-8984/22/i=9/a=096005} {\bibfield  {journal}
  {\bibinfo  {journal} {Journal of Physics: Condensed Matter}\ }\textbf
  {\bibinfo {volume} {22}},\ \bibinfo {pages} {096005} (\bibinfo {year}
  {2010})}\BibitemShut {NoStop}%
\bibitem [{\citenamefont {Bajpai}\ \emph {et~al.}(2017)\citenamefont {Bajpai},
  \citenamefont {Aslam}, \citenamefont {Hampel}, \citenamefont {Klingeler},\
  and\ \citenamefont {Grobert}}]{Ashna2}%
  \BibitemOpen
  \bibfield  {author} {\bibinfo {author} {\bibfnamefont {A.}~\bibnamefont
  {Bajpai}}, \bibinfo {author} {\bibfnamefont {Z.}~\bibnamefont {Aslam}},
  \bibinfo {author} {\bibfnamefont {S.}~\bibnamefont {Hampel}}, \bibinfo
  {author} {\bibfnamefont {R.}~\bibnamefont {Klingeler}},\ and\ \bibinfo
  {author} {\bibfnamefont {N.}~\bibnamefont {Grobert}},\ }\bibfield  {title}
  {\bibinfo {title} {A carbon-nanotube based nano-furnace for in-situ
  restructuring of a magnetoelectric oxide},\ }\href
  {https://doi.org/http://dx.doi.org/10.1016/j.carbon.2016.12.008} {\bibfield
  {journal} {\bibinfo  {journal} {Carbon}\ }\textbf {\bibinfo {volume} {114}},\
  \bibinfo {pages} {291 } (\bibinfo {year} {2017})}\BibitemShut {NoStop}%
\bibitem [{\citenamefont {Gross}\ \emph {et~al.}(2016)\citenamefont {Gross},
  \citenamefont {Mart\'{\i}nez}, \citenamefont {Tetienne}, \citenamefont
  {Hingant}, \citenamefont {Roch}, \citenamefont {Garcia}, \citenamefont
  {Soucaille}, \citenamefont {Adam}, \citenamefont {Kim}, \citenamefont
  {Rohart}, \citenamefont {Thiaville}, \citenamefont {Torrejon}, \citenamefont
  {Hayashi},\ and\ \citenamefont {Jacques}}]{Gross}%
  \BibitemOpen
  \bibfield  {author} {\bibinfo {author} {\bibfnamefont {I.}~\bibnamefont
  {Gross}}, \bibinfo {author} {\bibfnamefont {L.~J.}\ \bibnamefont
  {Mart\'{\i}nez}}, \bibinfo {author} {\bibfnamefont {J.-P.}\ \bibnamefont
  {Tetienne}}, \bibinfo {author} {\bibfnamefont {T.}~\bibnamefont {Hingant}},
  \bibinfo {author} {\bibfnamefont {J.-F.}\ \bibnamefont {Roch}}, \bibinfo
  {author} {\bibfnamefont {K.}~\bibnamefont {Garcia}}, \bibinfo {author}
  {\bibfnamefont {R.}~\bibnamefont {Soucaille}}, \bibinfo {author}
  {\bibfnamefont {J.~P.}\ \bibnamefont {Adam}}, \bibinfo {author}
  {\bibfnamefont {J.-V.}\ \bibnamefont {Kim}}, \bibinfo {author} {\bibfnamefont
  {S.}~\bibnamefont {Rohart}}, \bibinfo {author} {\bibfnamefont
  {A.}~\bibnamefont {Thiaville}}, \bibinfo {author} {\bibfnamefont
  {J.}~\bibnamefont {Torrejon}}, \bibinfo {author} {\bibfnamefont
  {M.}~\bibnamefont {Hayashi}},\ and\ \bibinfo {author} {\bibfnamefont
  {V.}~\bibnamefont {Jacques}},\ }\bibfield  {title} {\bibinfo {title} {Direct
  measurement of interfacial dzyaloshinskii-moriya interaction in
  $x|\text{CoFeB}|\text{MgO}$ heterostructures with a scanning nv magnetometer
  $(x=\text{Ta},\text{TaN}, \text{and W})$},\ }\href
  {https://doi.org/10.1103/PhysRevB.94.064413} {\bibfield  {journal} {\bibinfo
  {journal} {Phys. Rev. B}\ }\textbf {\bibinfo {volume} {94}},\ \bibinfo
  {pages} {064413} (\bibinfo {year} {2016})}\BibitemShut {NoStop}%
\bibitem [{\citenamefont {Beutier}\ \emph {et~al.}(2017)\citenamefont
  {Beutier}, \citenamefont {Collins}, \citenamefont {Dimitrova}, \citenamefont
  {Dmitrienko}, \citenamefont {Katsnelson}, \citenamefont {Kvashnin},
  \citenamefont {Lichtenstein}, \citenamefont {Mazurenko}, \citenamefont
  {Nisbet}, \citenamefont {Ovchinnikova},\ and\ \citenamefont
  {Pincini}}]{Beutier}%
  \BibitemOpen
  \bibfield  {author} {\bibinfo {author} {\bibfnamefont {G.}~\bibnamefont
  {Beutier}}, \bibinfo {author} {\bibfnamefont {S.~P.}\ \bibnamefont
  {Collins}}, \bibinfo {author} {\bibfnamefont {O.~V.}\ \bibnamefont
  {Dimitrova}}, \bibinfo {author} {\bibfnamefont {V.~E.}\ \bibnamefont
  {Dmitrienko}}, \bibinfo {author} {\bibfnamefont {M.~I.}\ \bibnamefont
  {Katsnelson}}, \bibinfo {author} {\bibfnamefont {Y.~O.}\ \bibnamefont
  {Kvashnin}}, \bibinfo {author} {\bibfnamefont {A.~I.}\ \bibnamefont
  {Lichtenstein}}, \bibinfo {author} {\bibfnamefont {V.~V.}\ \bibnamefont
  {Mazurenko}}, \bibinfo {author} {\bibfnamefont {A.~G.~A.}\ \bibnamefont
  {Nisbet}}, \bibinfo {author} {\bibfnamefont {E.~N.}\ \bibnamefont
  {Ovchinnikova}},\ and\ \bibinfo {author} {\bibfnamefont {D.}~\bibnamefont
  {Pincini}},\ }\bibfield  {title} {\bibinfo {title} {Band filling control of
  the dzyaloshinskii-moriya interaction in weakly ferromagnetic insulators},\
  }\href {https://doi.org/10.1103/PhysRevLett.119.167201} {\bibfield  {journal}
  {\bibinfo  {journal} {Phys. Rev. Lett.}\ }\textbf {\bibinfo {volume} {119}},\
  \bibinfo {pages} {167201} (\bibinfo {year} {2017})}\BibitemShut {NoStop}%
\bibitem [{\citenamefont {Kushauer}\ \emph {et~al.}(1994)\citenamefont
  {Kushauer}, \citenamefont {Kleemann}, \citenamefont {Mattsson},\ and\
  \citenamefont {Nordblad}}]{Kleemann}%
  \BibitemOpen
  \bibfield  {author} {\bibinfo {author} {\bibfnamefont {J.}~\bibnamefont
  {Kushauer}}, \bibinfo {author} {\bibfnamefont {W.}~\bibnamefont {Kleemann}},
  \bibinfo {author} {\bibfnamefont {J.}~\bibnamefont {Mattsson}},\ and\
  \bibinfo {author} {\bibfnamefont {P.}~\bibnamefont {Nordblad}},\ }\bibfield
  {title} {\bibinfo {title} {Crossover from logarithmically relaxing to
  piezomagnetically frozen magnetic remanence in low-field-cooled
  ${\mathrm{fe}}_{0.47}$${\mathrm{zn}}_{0.53}$${\mathrm{f}}_{2}$},\ }\href
  {https://doi.org/10.1103/PhysRevB.49.6346} {\bibfield  {journal} {\bibinfo
  {journal} {Phys. Rev. B}\ }\textbf {\bibinfo {volume} {49}},\ \bibinfo
  {pages} {6346} (\bibinfo {year} {1994})}\BibitemShut {NoStop}%
\bibitem [{\citenamefont {Binder}\ and\ \citenamefont {Young}(1986)}]{Binder}%
  \BibitemOpen
  \bibfield  {author} {\bibinfo {author} {\bibfnamefont {K.}~\bibnamefont
  {Binder}}\ and\ \bibinfo {author} {\bibfnamefont {A.~P.}\ \bibnamefont
  {Young}},\ }\bibfield  {title} {\bibinfo {title} {Spin glasses: Experimental
  facts, theoretical concepts, and open questions},\ }\href
  {https://doi.org/10.1103/RevModPhys.58.801} {\bibfield  {journal} {\bibinfo
  {journal} {Rev. Mod. Phys.}\ }\textbf {\bibinfo {volume} {58}},\ \bibinfo
  {pages} {801} (\bibinfo {year} {1986})}\BibitemShut {NoStop}%
\bibitem [{\citenamefont {Kreines}\ and\ \citenamefont
  {Shal'Nikova}(1970)}]{Kreines}%
  \BibitemOpen
  \bibfield  {author} {\bibinfo {author} {\bibfnamefont {N.}~\bibnamefont
  {Kreines}}\ and\ \bibinfo {author} {\bibfnamefont {T.}~\bibnamefont
  {Shal'Nikova}},\ }\bibfield  {title} {\bibinfo {title} {Weak ferromagnetism
  in nic03},\ }\href@noop {} {\bibfield  {journal} {\bibinfo  {journal} {Soviet
  Physics JETP}\ }\textbf {\bibinfo {volume} {31}} (\bibinfo {year}
  {1970})}\BibitemShut {NoStop}%
\bibitem [{\citenamefont {Srivastava}(1970)}]{Srivastava}%
  \BibitemOpen
  \bibfield  {author} {\bibinfo {author} {\bibfnamefont {V.~C.}\ \bibnamefont
  {Srivastava}},\ }\bibfield  {title} {\bibinfo {title} {Effect of pressure on
  the néel temperature of ${\mathrm{mnco}}_{3}$, ${\mathrm{coco}}_{3}$, and
  ${\mathrm{feco}}_{3}$},\ }\href {https://doi.org/10.1063/1.1658873}
  {\bibfield  {journal} {\bibinfo  {journal} {Journal of Applied Physics}\
  }\textbf {\bibinfo {volume} {41}},\ \bibinfo {pages} {1190} (\bibinfo {year}
  {1970})}\BibitemShut {NoStop}%
\bibitem [{\citenamefont {Morrish}(1995)}]{Morrish}%
  \BibitemOpen
  \bibfield  {author} {\bibinfo {author} {\bibfnamefont {A.~H.}\ \bibnamefont
  {Morrish}},\ }\href {https://doi.org/10.1142/2518} {\emph {\bibinfo {title}
  {Canted Antiferromagnetism: Hematite}}}\ (\bibinfo  {publisher} {WORLD
  SCIENTIFIC},\ \bibinfo {year} {1995})\BibitemShut {NoStop}%
\bibitem [{\citenamefont {Radhakrishna}(1982)}]{Radhakrishna}%
  \BibitemOpen
  \bibfield  {author} {\bibinfo {author} {\bibfnamefont {P.}~\bibnamefont
  {Radhakrishna}},\ }\bibfield  {title} {\bibinfo {title} {Polarized neutron
  diffraction studies on weak ferromagnetism-a survey},\ }\href@noop {}
  {\bibfield  {journal} {\bibinfo  {journal} {Le Journal de Physique
  Colloques}\ }\textbf {\bibinfo {volume} {43}},\ \bibinfo {pages} {C7}
  (\bibinfo {year} {1982})}\BibitemShut {NoStop}%
\bibitem [{\citenamefont {Martynov}(2018)}]{Martynov}%
  \BibitemOpen
  \bibfield  {author} {\bibinfo {author} {\bibfnamefont {S.~N.}\ \bibnamefont
  {Martynov}},\ }\bibfield  {title} {\bibinfo {title} {Single-ion mechanism of
  weak ferromagnetism and a spin-flop transition in one-and two-position
  antiferromagnets},\ }\href@noop {} {\bibfield  {journal} {\bibinfo  {journal}
  {JETP Letters}\ }\textbf {\bibinfo {volume} {108}},\ \bibinfo {pages} {196}
  (\bibinfo {year} {2018})}\BibitemShut {NoStop}%
\bibitem [{\citenamefont {Borovik-Romanov}\ \emph {et~al.}(1973)\citenamefont
  {Borovik-Romanov}, \citenamefont {Bazhan},\ and\ \citenamefont
  {Kreines}}]{Bazhan}%
  \BibitemOpen
  \bibfield  {author} {\bibinfo {author} {\bibfnamefont {A.}~\bibnamefont
  {Borovik-Romanov}}, \bibinfo {author} {\bibfnamefont {A.}~\bibnamefont
  {Bazhan}},\ and\ \bibinfo {author} {\bibfnamefont {N.}~\bibnamefont
  {Kreines}},\ }\bibfield  {title} {\bibinfo {title} {The weak ferromagnetism
  of ${\mathrm{nif}}_{2}$},\ }\href@noop {} {\bibfield  {journal} {\bibinfo
  {journal} {Soviet Journal of Experimental and Theoretical Physics}\ }\textbf
  {\bibinfo {volume} {37}},\ \bibinfo {pages} {695} (\bibinfo {year}
  {1973})}\BibitemShut {NoStop}%
\bibitem [{\citenamefont {Romanov}(1960)}]{Romanov1}%
  \BibitemOpen
  \bibfield  {author} {\bibinfo {author} {\bibfnamefont {A.~B.}\ \bibnamefont
  {Romanov}},\ }\href@noop {} {\bibfield  {journal} {\bibinfo  {journal}
  {Soviet Physics JETP}\ }\textbf {\bibinfo {volume} {11}},\ \bibinfo {pages}
  {786} (\bibinfo {year} {1960})}\BibitemShut {NoStop}%
\bibitem [{\citenamefont {Loktev}(1981)}]{Loktev}%
  \BibitemOpen
  \bibfield  {author} {\bibinfo {author} {\bibfnamefont {V.}~\bibnamefont
  {Loktev}},\ }\bibfield  {title} {\bibinfo {title} {Spin excitations of
  ${\mathrm{coco}}_{3}$ in a longitudinal magnetic field},\ }\href@noop {}
  {\bibfield  {journal} {\bibinfo  {journal} {Physics Letters A}\ }\textbf
  {\bibinfo {volume} {81}},\ \bibinfo {pages} {187} (\bibinfo {year}
  {1981})}\BibitemShut {NoStop}%
\bibitem [{\citenamefont {Benitez}\ \emph
  {et~al.}(2008{\natexlab{a}})\citenamefont {Benitez}, \citenamefont
  {Petracic}, \citenamefont {Salabas}, \citenamefont {Radu}, \citenamefont
  {T\"uys\"uz}, \citenamefont {Sch\"uth},\ and\ \citenamefont
  {Zabel}}]{Benitez1}%
  \BibitemOpen
  \bibfield  {author} {\bibinfo {author} {\bibfnamefont {M.~J.}\ \bibnamefont
  {Benitez}}, \bibinfo {author} {\bibfnamefont {O.}~\bibnamefont {Petracic}},
  \bibinfo {author} {\bibfnamefont {E.~L.}\ \bibnamefont {Salabas}}, \bibinfo
  {author} {\bibfnamefont {F.}~\bibnamefont {Radu}}, \bibinfo {author}
  {\bibfnamefont {H.}~\bibnamefont {T\"uys\"uz}}, \bibinfo {author}
  {\bibfnamefont {F.}~\bibnamefont {Sch\"uth}},\ and\ \bibinfo {author}
  {\bibfnamefont {H.}~\bibnamefont {Zabel}},\ }\bibfield  {title} {\bibinfo
  {title} {Evidence for core-shell magnetic behavior in antiferromagnetic
  ${\mathrm{co}}_{3}{\mathrm{o}}_{4}$ nanowires},\ }\href
  {https://doi.org/10.1103/PhysRevLett.101.097206} {\bibfield  {journal}
  {\bibinfo  {journal} {Phys. Rev. Lett.}\ }\textbf {\bibinfo {volume} {101}},\
  \bibinfo {pages} {097206} (\bibinfo {year} {2008}{\natexlab{a}})}\BibitemShut
  {NoStop}%
\bibitem [{\citenamefont {Benitez}\ \emph
  {et~al.}(2011{\natexlab{a}})\citenamefont {Benitez}, \citenamefont
  {Petracic}, \citenamefont {T\"uys\"uz}, \citenamefont {Sch\"uth},\ and\
  \citenamefont {Zabel}}]{Benitez2}%
  \BibitemOpen
  \bibfield  {author} {\bibinfo {author} {\bibfnamefont {M.~J.}\ \bibnamefont
  {Benitez}}, \bibinfo {author} {\bibfnamefont {O.}~\bibnamefont {Petracic}},
  \bibinfo {author} {\bibfnamefont {H.}~\bibnamefont {T\"uys\"uz}}, \bibinfo
  {author} {\bibfnamefont {F.}~\bibnamefont {Sch\"uth}},\ and\ \bibinfo
  {author} {\bibfnamefont {H.}~\bibnamefont {Zabel}},\ }\bibfield  {title}
  {\bibinfo {title} {Fingerprinting the magnetic behavior of antiferromagnetic
  nanostructures using remanent magnetization curves},\ }\href
  {https://doi.org/10.1103/PhysRevB.83.134424} {\bibfield  {journal} {\bibinfo
  {journal} {Phys. Rev. B}\ }\textbf {\bibinfo {volume} {83}},\ \bibinfo
  {pages} {134424} (\bibinfo {year} {2011}{\natexlab{a}})}\BibitemShut
  {NoStop}%
\bibitem [{\citenamefont {Dzyaloshinskii}(1957{\natexlab{b}})}]{Dzy2}%
  \BibitemOpen
  \bibfield  {author} {\bibinfo {author} {\bibfnamefont {I.~E.}\ \bibnamefont
  {Dzyaloshinskii}},\ }\href@noop {} {\bibfield  {journal} {\bibinfo  {journal}
  {JETP}\ }\textbf {\bibinfo {volume} {33}},\ \bibinfo {pages} {807} (\bibinfo
  {year} {1957}{\natexlab{b}})}\BibitemShut {NoStop}%
\bibitem [{\citenamefont {Landau}\ and\ \citenamefont
  {Lifshitz}(1984)}]{Landau}%
  \BibitemOpen
  \bibfield  {author} {\bibinfo {author} {\bibfnamefont {L.~D.}\ \bibnamefont
  {Landau}}\ and\ \bibinfo {author} {\bibfnamefont {E.}~\bibnamefont
  {Lifshitz}},\ }\href@noop {} {\emph {\bibinfo {title} {Electrodynamics of
  Continuous Media}}}\ (\bibinfo  {publisher} {Pergamon Press Ltd.},\ \bibinfo
  {year} {1984})\BibitemShut {NoStop}%
\bibitem [{\citenamefont {Kimber}\ and\ \citenamefont
  {Attfield}(2007{\natexlab{a}})}]{Attfield}%
  \BibitemOpen
  \bibfield  {author} {\bibinfo {author} {\bibfnamefont {S.~A.~J.}\
  \bibnamefont {Kimber}}\ and\ \bibinfo {author} {\bibfnamefont {J.~P.}\
  \bibnamefont {Attfield}},\ }\bibfield  {title} {\bibinfo {title} {Magnetic
  order in acentric pb2mno4},\ }\href {https://doi.org/10.1039/B704361A}
  {\bibfield  {journal} {\bibinfo  {journal} {J. Mater. Chem.}\ }\textbf
  {\bibinfo {volume} {17}},\ \bibinfo {pages} {4885} (\bibinfo {year}
  {2007}{\natexlab{a}})}\BibitemShut {NoStop}%
\bibitem [{\citenamefont {Plumier}\ \emph {et~al.}(1985)\citenamefont
  {Plumier}, \citenamefont {Sougi}, \citenamefont {Lecomte},\ and\
  \citenamefont {Saint‐James}}]{Plumier1}%
  \BibitemOpen
  \bibfield  {author} {\bibinfo {author} {\bibfnamefont {R.}~\bibnamefont
  {Plumier}}, \bibinfo {author} {\bibfnamefont {M.}~\bibnamefont {Sougi}},
  \bibinfo {author} {\bibfnamefont {M.}~\bibnamefont {Lecomte}},\ and\ \bibinfo
  {author} {\bibfnamefont {R.}~\bibnamefont {Saint‐James}},\ }\bibfield
  {title} {\bibinfo {title} {Weak ferromagnetism in ${\mathrm{nico}}_{3}$},\
  }\href {https://doi.org/10.1063/1.335118} {\bibfield  {journal} {\bibinfo
  {journal} {Journal of Applied Physics}\ }\textbf {\bibinfo {volume} {57}},\
  \bibinfo {pages} {3261} (\bibinfo {year} {1985})}\BibitemShut {NoStop}%
\bibitem [{\citenamefont {Plumier}\ \emph {et~al.}(1983)\citenamefont
  {Plumier}, \citenamefont {Sougi},\ and\ \citenamefont
  {Saint-James}}]{Plumier2}%
  \BibitemOpen
  \bibfield  {author} {\bibinfo {author} {\bibfnamefont {R.}~\bibnamefont
  {Plumier}}, \bibinfo {author} {\bibfnamefont {M.}~\bibnamefont {Sougi}},\
  and\ \bibinfo {author} {\bibfnamefont {R.}~\bibnamefont {Saint-James}},\
  }\bibfield  {title} {\bibinfo {title} {Neutron-diffraction reinvestigation of
  nic${\mathrm{o}}_{3}$},\ }\href {https://doi.org/10.1103/PhysRevB.28.4016}
  {\bibfield  {journal} {\bibinfo  {journal} {Phys. Rev. B}\ }\textbf {\bibinfo
  {volume} {28}},\ \bibinfo {pages} {4016} (\bibinfo {year}
  {1983})}\BibitemShut {NoStop}%
\bibitem [{\citenamefont {Smith}(1916)}]{Smith}%
  \BibitemOpen
  \bibfield  {author} {\bibinfo {author} {\bibfnamefont {T.~T.}\ \bibnamefont
  {Smith}},\ }\bibfield  {title} {\bibinfo {title} {The magnetic properties of
  hematite},\ }\href@noop {} {\bibfield  {journal} {\bibinfo  {journal}
  {Physical Review}\ }\textbf {\bibinfo {volume} {8}},\ \bibinfo {pages} {721}
  (\bibinfo {year} {1916})}\BibitemShut {NoStop}%
\bibitem [{\citenamefont {McDannald}\ \emph {et~al.}(2015)\citenamefont
  {McDannald}, \citenamefont {Kuna}, \citenamefont {Seehra},\ and\
  \citenamefont {Jain}}]{McDannald}%
  \BibitemOpen
  \bibfield  {author} {\bibinfo {author} {\bibfnamefont {A.}~\bibnamefont
  {McDannald}}, \bibinfo {author} {\bibfnamefont {L.}~\bibnamefont {Kuna}},
  \bibinfo {author} {\bibfnamefont {M.~S.}\ \bibnamefont {Seehra}},\ and\
  \bibinfo {author} {\bibfnamefont {M.}~\bibnamefont {Jain}},\ }\bibfield
  {title} {\bibinfo {title} {Magnetic exchange interactions of
  rare-earth-substituted ${\mathrm{dycro}}_{3}$ bulk powders},\ }\href
  {https://doi.org/10.1103/PhysRevB.91.224415} {\bibfield  {journal} {\bibinfo
  {journal} {Phys. Rev. B}\ }\textbf {\bibinfo {volume} {91}},\ \bibinfo
  {pages} {224415} (\bibinfo {year} {2015})}\BibitemShut {NoStop}%
\bibitem [{\citenamefont {Khomskii}(2014)}]{Khomskii}%
  \BibitemOpen
  \bibfield  {author} {\bibinfo {author} {\bibfnamefont {D.~I.}\ \bibnamefont
  {Khomskii}},\ }\href@noop {} {\emph {\bibinfo {title} {Transition metal
  compounds}}}\ (\bibinfo  {publisher} {Cambridge University Press},\ \bibinfo
  {year} {2014})\BibitemShut {NoStop}%
\bibitem [{\citenamefont {Morin}(1950)}]{Morin1}%
  \BibitemOpen
  \bibfield  {author} {\bibinfo {author} {\bibfnamefont {F.}~\bibnamefont
  {Morin}},\ }\bibfield  {title} {\bibinfo {title} {Magnetic susceptibility of
  $\alpha$ fe 2 o 3 and $\alpha$ fe 2 o 3 with added titanium},\ }\href@noop {}
  {\bibfield  {journal} {\bibinfo  {journal} {Physical Review}\ }\textbf
  {\bibinfo {volume} {78}},\ \bibinfo {pages} {819} (\bibinfo {year}
  {1950})}\BibitemShut {NoStop}%
\bibitem [{\citenamefont {Onose}\ \emph {et~al.}(2010)\citenamefont {Onose},
  \citenamefont {Ideue}, \citenamefont {Katsura}, \citenamefont {Shiomi},
  \citenamefont {Nagaosa},\ and\ \citenamefont {Tokura}}]{Onose}%
  \BibitemOpen
  \bibfield  {author} {\bibinfo {author} {\bibfnamefont {Y.}~\bibnamefont
  {Onose}}, \bibinfo {author} {\bibfnamefont {T.}~\bibnamefont {Ideue}},
  \bibinfo {author} {\bibfnamefont {H.}~\bibnamefont {Katsura}}, \bibinfo
  {author} {\bibfnamefont {Y.}~\bibnamefont {Shiomi}}, \bibinfo {author}
  {\bibfnamefont {N.}~\bibnamefont {Nagaosa}},\ and\ \bibinfo {author}
  {\bibfnamefont {Y.}~\bibnamefont {Tokura}},\ }\bibfield  {title} {\bibinfo
  {title} {Observation of the magnon hall effect},\ }\href
  {https://doi.org/10.1126/science.1188260} {\bibfield  {journal} {\bibinfo
  {journal} {Science}\ }\textbf {\bibinfo {volume} {329}},\ \bibinfo {pages}
  {297} (\bibinfo {year} {2010})}\BibitemShut {NoStop}%
\bibitem [{\citenamefont {Hasan}\ and\ \citenamefont {Kane}(2010)}]{Hasan}%
  \BibitemOpen
  \bibfield  {author} {\bibinfo {author} {\bibfnamefont {M.~Z.}\ \bibnamefont
  {Hasan}}\ and\ \bibinfo {author} {\bibfnamefont {C.~L.}\ \bibnamefont
  {Kane}},\ }\bibfield  {title} {\bibinfo {title} {Colloquium},\ }\href
  {https://doi.org/10.1103/RevModPhys.82.3045} {\bibfield  {journal} {\bibinfo
  {journal} {Rev. Mod. Phys.}\ }\textbf {\bibinfo {volume} {82}},\ \bibinfo
  {pages} {3045} (\bibinfo {year} {2010})}\BibitemShut {NoStop}%
\bibitem [{\citenamefont {Gomonay}\ \emph {et~al.}(2018)\citenamefont
  {Gomonay}, \citenamefont {Baltz}, \citenamefont {Brataas},\ and\
  \citenamefont {Tserkovnyak}}]{Gomonay}%
  \BibitemOpen
  \bibfield  {author} {\bibinfo {author} {\bibfnamefont {O.}~\bibnamefont
  {Gomonay}}, \bibinfo {author} {\bibfnamefont {V.}~\bibnamefont {Baltz}},
  \bibinfo {author} {\bibfnamefont {A.}~\bibnamefont {Brataas}},\ and\ \bibinfo
  {author} {\bibfnamefont {Y.}~\bibnamefont {Tserkovnyak}},\ }\href@noop {}
  {\bibfield  {journal} {\bibinfo  {journal} {Nature Physics}\ }\textbf
  {\bibinfo {volume} {14}},\ \bibinfo {pages} {213} (\bibinfo {year}
  {2018})}\BibitemShut {NoStop}%
\bibitem [{\citenamefont {Borovik-Romanov}(1960)}]{Romanov2}%
  \BibitemOpen
  \bibfield  {author} {\bibinfo {author} {\bibfnamefont {A.~S.}\ \bibnamefont
  {Borovik-Romanov}},\ }\href@noop {} {\bibfield  {journal} {\bibinfo
  {journal} {Zh. Eksp. Teor. Fiz.}\ }\textbf {\bibinfo {volume} {38}},\
  \bibinfo {pages} {1088} (\bibinfo {year} {1960})}\BibitemShut {NoStop}%
\bibitem [{\citenamefont {Andratskii}\ and\ \citenamefont
  {Borovik-Romanov}(1966)}]{Romanov3}%
  \BibitemOpen
  \bibfield  {author} {\bibinfo {author} {\bibfnamefont {V.}~\bibnamefont
  {Andratskii}}\ and\ \bibinfo {author} {\bibfnamefont {A.~S.}\ \bibnamefont
  {Borovik-Romanov}},\ }\href@noop {} {\bibfield  {journal} {\bibinfo
  {journal} {JETP}\ }\textbf {\bibinfo {volume} {687}},\ \bibinfo {pages}
  {1036} (\bibinfo {year} {1966})}\BibitemShut {NoStop}%
\bibitem [{\citenamefont {Borovik-Romanov}(1994)}]{Romanov4}%
  \BibitemOpen
  \bibfield  {author} {\bibinfo {author} {\bibfnamefont {A.~S.}\ \bibnamefont
  {Borovik-Romanov}},\ }\href@noop {} {\bibfield  {journal} {\bibinfo
  {journal} {Ferroelectrics}\ }\textbf {\bibinfo {volume} {162}},\ \bibinfo
  {pages} {153} (\bibinfo {year} {1994})}\BibitemShut {NoStop}%
\bibitem [{\citenamefont {Kimber}\ and\ \citenamefont
  {Attfield}(2007{\natexlab{b}})}]{Paul}%
  \BibitemOpen
  \bibfield  {author} {\bibinfo {author} {\bibfnamefont {S.~A.~J.}\
  \bibnamefont {Kimber}}\ and\ \bibinfo {author} {\bibfnamefont {J.~P.}\
  \bibnamefont {Attfield}},\ }\bibfield  {title} {\bibinfo {title} {Magnetic
  order in acentric pb2mno4},\ }\href {https://doi.org/10.1039/B704361A}
  {\bibfield  {journal} {\bibinfo  {journal} {J. Mater. Chem.}\ }\textbf
  {\bibinfo {volume} {17}},\ \bibinfo {pages} {4885} (\bibinfo {year}
  {2007}{\natexlab{b}})}\BibitemShut {NoStop}%
\bibitem [{\citenamefont {Phillips}\ \emph
  {et~al.}(1967{\natexlab{a}})\citenamefont {Phillips}, \citenamefont
  {Townsend},\ and\ \citenamefont {White}}]{Philip1}%
  \BibitemOpen
  \bibfield  {author} {\bibinfo {author} {\bibfnamefont {T.~G.}\ \bibnamefont
  {Phillips}}, \bibinfo {author} {\bibfnamefont {R.~L.}\ \bibnamefont
  {Townsend}},\ and\ \bibinfo {author} {\bibfnamefont {R.~L.}\ \bibnamefont
  {White}},\ }\bibfield  {title} {\bibinfo {title} {Piezomagnetism of
  co${\mathrm{f}}_{2}$ and
  $\ensuremath{\alpha}$-${\mathrm{fe}}_{2}$${\mathrm{o}}_{3}$ from
  electron-paramagnetic-resonance pressure experiments},\ }\href
  {https://doi.org/10.1103/PhysRevLett.18.646} {\bibfield  {journal} {\bibinfo
  {journal} {Phys. Rev. Lett.}\ }\textbf {\bibinfo {volume} {18}},\ \bibinfo
  {pages} {646} (\bibinfo {year} {1967}{\natexlab{a}})}\BibitemShut {NoStop}%
\bibitem [{\citenamefont {Phillips}\ \emph
  {et~al.}(1967{\natexlab{b}})\citenamefont {Phillips}, \citenamefont
  {Townsend},\ and\ \citenamefont {White}}]{Philip2}%
  \BibitemOpen
  \bibfield  {author} {\bibinfo {author} {\bibfnamefont {T.~G.}\ \bibnamefont
  {Phillips}}, \bibinfo {author} {\bibfnamefont {R.~L.}\ \bibnamefont
  {Townsend}},\ and\ \bibinfo {author} {\bibfnamefont {R.~L.}\ \bibnamefont
  {White}},\ }\bibfield  {title} {\bibinfo {title} {Piezomagnetism of
  $\ensuremath{\alpha}$-${\mathrm{fe}}_{2}$${\mathrm{o}}_{3}$ and the
  magnetoelastic tensor of ${\mathrm{fe}}^{3+}$ in
  ${\mathrm{al}}_{2}$${\mathrm{o}}_{3}$},\ }\href
  {https://doi.org/10.1103/PhysRev.162.382} {\bibfield  {journal} {\bibinfo
  {journal} {Phys. Rev.}\ }\textbf {\bibinfo {volume} {162}},\ \bibinfo {pages}
  {382} (\bibinfo {year} {1967}{\natexlab{b}})}\BibitemShut {NoStop}%
\bibitem [{\citenamefont {{Binek, Ch.}}\ \emph {et~al.}(2005)\citenamefont
  {{Binek, Ch.}}, \citenamefont {{Borisov, P.}}, \citenamefont {{Chen, Xi}},
  \citenamefont {{Hochstrat, A.}}, \citenamefont {{Sahoo, S.}},\ and\
  \citenamefont {{Kleemann, W.}}}]{Binek1}%
  \BibitemOpen
  \bibfield  {author} {\bibinfo {author} {\bibnamefont {{Binek, Ch.}}},
  \bibinfo {author} {\bibnamefont {{Borisov, P.}}}, \bibinfo {author}
  {\bibnamefont {{Chen, Xi}}}, \bibinfo {author} {\bibnamefont {{Hochstrat,
  A.}}}, \bibinfo {author} {\bibnamefont {{Sahoo, S.}}},\ and\ \bibinfo
  {author} {\bibnamefont {{Kleemann, W.}}},\ }\bibfield  {title} {\bibinfo
  {title} {Perpendicular exchange bias and its control by magnetic, stress and
  electric fields},\ }\href {https://doi.org/10.1140/epjb/e2005-00054-2}
  {\bibfield  {journal} {\bibinfo  {journal} {Eur. Phys. J. B}\ }\textbf
  {\bibinfo {volume} {45}},\ \bibinfo {pages} {197} (\bibinfo {year}
  {2005})}\BibitemShut {NoStop}%
\bibitem [{\citenamefont {O¨zdemir}\ and\ \citenamefont
  {Dunlop}(2008)}]{Dunlop}%
  \BibitemOpen
  \bibfield  {author} {\bibinfo {author} {\bibfnamefont {O.}~\bibnamefont
  {O¨zdemir}}\ and\ \bibinfo {author} {\bibfnamefont {D.~J.}\ \bibnamefont
  {Dunlop}},\ }\href@noop {} {\bibfield  {journal} {\bibinfo  {journal}
  {Geochem. Geophys. Geosyst.}\ }\textbf {\bibinfo {volume} {9}},\ \bibinfo
  {pages} {1} (\bibinfo {year} {2008})}\BibitemShut {NoStop}%
\bibitem [{\citenamefont {Amin}\ and\ \citenamefont {Arajs}(1987)}]{Amin}%
  \BibitemOpen
  \bibfield  {author} {\bibinfo {author} {\bibfnamefont {N.}~\bibnamefont
  {Amin}}\ and\ \bibinfo {author} {\bibfnamefont {S.}~\bibnamefont {Arajs}},\
  }\bibfield  {title} {\bibinfo {title} {Morin temperature of annealed
  submicronic \ensuremath{\alpha}-${\mathrm{f}}_{2}$${\mathrm{o}}_{3}$
  particles},\ }\href {https://doi.org/10.1103/PhysRevB.35.4810} {\bibfield
  {journal} {\bibinfo  {journal} {Phys. Rev. B}\ }\textbf {\bibinfo {volume}
  {35}},\ \bibinfo {pages} {4810} (\bibinfo {year} {1987})}\BibitemShut
  {NoStop}%
\bibitem [{\citenamefont {Zysler}\ \emph {et~al.}(2003)\citenamefont {Zysler},
  \citenamefont {Fiorani}, \citenamefont {Testa}, \citenamefont {Suber},
  \citenamefont {Agostinelli},\ and\ \citenamefont {Godinho}}]{Zysler}%
  \BibitemOpen
  \bibfield  {author} {\bibinfo {author} {\bibfnamefont {R.~D.}\ \bibnamefont
  {Zysler}}, \bibinfo {author} {\bibfnamefont {D.}~\bibnamefont {Fiorani}},
  \bibinfo {author} {\bibfnamefont {A.~M.}\ \bibnamefont {Testa}}, \bibinfo
  {author} {\bibfnamefont {L.}~\bibnamefont {Suber}}, \bibinfo {author}
  {\bibfnamefont {E.}~\bibnamefont {Agostinelli}},\ and\ \bibinfo {author}
  {\bibfnamefont {M.}~\bibnamefont {Godinho}},\ }\bibfield  {title} {\bibinfo
  {title} {Size dependence of the spin-flop transition in hematite
  nanoparticles},\ }\href {https://doi.org/10.1103/PhysRevB.68.212408}
  {\bibfield  {journal} {\bibinfo  {journal} {Phys. Rev. B}\ }\textbf {\bibinfo
  {volume} {68}},\ \bibinfo {pages} {212408} (\bibinfo {year}
  {2003})}\BibitemShut {NoStop}%
\bibitem [{\citenamefont {Mitra}\ \emph {et~al.}(2009)\citenamefont {Mitra},
  \citenamefont {Das}, \citenamefont {Basu}, \citenamefont {Sahu},\ and\
  \citenamefont {Mandal}}]{Mitra}%
  \BibitemOpen
  \bibfield  {author} {\bibinfo {author} {\bibfnamefont {S.}~\bibnamefont
  {Mitra}}, \bibinfo {author} {\bibfnamefont {S.}~\bibnamefont {Das}}, \bibinfo
  {author} {\bibfnamefont {S.}~\bibnamefont {Basu}}, \bibinfo {author}
  {\bibfnamefont {P.}~\bibnamefont {Sahu}},\ and\ \bibinfo {author}
  {\bibfnamefont {K.}~\bibnamefont {Mandal}},\ }\bibfield  {title} {\bibinfo
  {title} {Shape- and field-dependent morin transitions in structured
  $\alpha$-fe$_2$o$_3$},\ }\href
  {https://doi.org/https://doi.org/10.1016/j.jmmm.2009.04.044} {\bibfield
  {journal} {\bibinfo  {journal} {Journal of Magnetism and Magnetic Materials}\
  }\textbf {\bibinfo {volume} {321}},\ \bibinfo {pages} {2925 } (\bibinfo
  {year} {2009})}\BibitemShut {NoStop}%
\bibitem [{\citenamefont {M\o{}rup}\ and\ \citenamefont
  {Frandsen}(2004)}]{Morup}%
  \BibitemOpen
  \bibfield  {author} {\bibinfo {author} {\bibfnamefont {S.}~\bibnamefont
  {M\o{}rup}}\ and\ \bibinfo {author} {\bibfnamefont {C.}~\bibnamefont
  {Frandsen}},\ }\bibfield  {title} {\bibinfo {title} {Thermoinduced
  magnetization in nanoparticles of antiferromagnetic materials},\ }\href
  {https://doi.org/10.1103/PhysRevLett.92.217201} {\bibfield  {journal}
  {\bibinfo  {journal} {Phys. Rev. Lett.}\ }\textbf {\bibinfo {volume} {92}},\
  \bibinfo {pages} {217201} (\bibinfo {year} {2004})}\BibitemShut {NoStop}%
\bibitem [{\citenamefont {Benitez}\ \emph
  {et~al.}(2011{\natexlab{b}})\citenamefont {Benitez}, \citenamefont
  {Petracic}, \citenamefont {T\"uys\"uz}, \citenamefont {Sch\"uth},\ and\
  \citenamefont {Zabel}}]{Ben1}%
  \BibitemOpen
  \bibfield  {author} {\bibinfo {author} {\bibfnamefont {M.~J.}\ \bibnamefont
  {Benitez}}, \bibinfo {author} {\bibfnamefont {O.}~\bibnamefont {Petracic}},
  \bibinfo {author} {\bibfnamefont {H.}~\bibnamefont {T\"uys\"uz}}, \bibinfo
  {author} {\bibfnamefont {F.}~\bibnamefont {Sch\"uth}},\ and\ \bibinfo
  {author} {\bibfnamefont {H.}~\bibnamefont {Zabel}},\ }\bibfield  {title}
  {\bibinfo {title} {Fingerprinting the magnetic behavior of antiferromagnetic
  nanostructures using remanent magnetization curves},\ }\href
  {https://doi.org/10.1103/PhysRevB.83.134424} {\bibfield  {journal} {\bibinfo
  {journal} {Phys. Rev. B}\ }\textbf {\bibinfo {volume} {83}},\ \bibinfo
  {pages} {134424} (\bibinfo {year} {2011}{\natexlab{b}})}\BibitemShut
  {NoStop}%
\bibitem [{\citenamefont {Benitez}\ \emph
  {et~al.}(2008{\natexlab{b}})\citenamefont {Benitez}, \citenamefont
  {Petracic}, \citenamefont {Salabas}, \citenamefont {Radu}, \citenamefont
  {T\"uys\"uz}, \citenamefont {Sch\"uth},\ and\ \citenamefont {Zabel}}]{Ben2}%
  \BibitemOpen
  \bibfield  {author} {\bibinfo {author} {\bibfnamefont {M.~J.}\ \bibnamefont
  {Benitez}}, \bibinfo {author} {\bibfnamefont {O.}~\bibnamefont {Petracic}},
  \bibinfo {author} {\bibfnamefont {E.~L.}\ \bibnamefont {Salabas}}, \bibinfo
  {author} {\bibfnamefont {F.}~\bibnamefont {Radu}}, \bibinfo {author}
  {\bibfnamefont {H.}~\bibnamefont {T\"uys\"uz}}, \bibinfo {author}
  {\bibfnamefont {F.}~\bibnamefont {Sch\"uth}},\ and\ \bibinfo {author}
  {\bibfnamefont {H.}~\bibnamefont {Zabel}},\ }\bibfield  {title} {\bibinfo
  {title} {Evidence for core-shell magnetic behavior in antiferromagnetic
  ${\mathrm{co}}_{3}{\mathrm{o}}_{4}$ nanowires},\ }\href
  {https://doi.org/10.1103/PhysRevLett.101.097206} {\bibfield  {journal}
  {\bibinfo  {journal} {Phys. Rev. Lett.}\ }\textbf {\bibinfo {volume} {101}},\
  \bibinfo {pages} {097206} (\bibinfo {year} {2008}{\natexlab{b}})}\BibitemShut
  {NoStop}%
\bibitem [{\citenamefont {Mattsson}\ \emph {et~al.}(2000)\citenamefont
  {Mattsson}, \citenamefont {Djurberg},\ and\ \citenamefont {Nordblad}}]{Mat}%
  \BibitemOpen
  \bibfield  {author} {\bibinfo {author} {\bibfnamefont {J.}~\bibnamefont
  {Mattsson}}, \bibinfo {author} {\bibfnamefont {C.}~\bibnamefont {Djurberg}},\
  and\ \bibinfo {author} {\bibfnamefont {P.}~\bibnamefont {Nordblad}},\
  }\bibfield  {title} {\bibinfo {title} {Low-temperature magnetization in
  dilute ising antiferromagnets},\ }\href
  {https://doi.org/10.1103/PhysRevB.61.11274} {\bibfield  {journal} {\bibinfo
  {journal} {Phys. Rev. B}\ }\textbf {\bibinfo {volume} {61}},\ \bibinfo
  {pages} {11274} (\bibinfo {year} {2000})}\BibitemShut {NoStop}%
\bibitem [{\citenamefont {Suzuki}\ \emph {et~al.}(2006)\citenamefont {Suzuki},
  \citenamefont {Suzuki},\ and\ \citenamefont {Matsuura}}]{Suzuki}%
  \BibitemOpen
  \bibfield  {author} {\bibinfo {author} {\bibfnamefont {M.}~\bibnamefont
  {Suzuki}}, \bibinfo {author} {\bibfnamefont {I.~S.}\ \bibnamefont {Suzuki}},\
  and\ \bibinfo {author} {\bibfnamefont {M.}~\bibnamefont {Matsuura}},\
  }\bibfield  {title} {\bibinfo {title} {Memory and aging effect in
  hierarchical spin orderings of the stage-2 $\mathrm{Co}{\mathrm{cl}}_{2}$
  graphite intercalation compound},\ }\href
  {https://doi.org/10.1103/PhysRevB.73.184414} {\bibfield  {journal} {\bibinfo
  {journal} {Phys. Rev. B}\ }\textbf {\bibinfo {volume} {73}},\ \bibinfo
  {pages} {184414} (\bibinfo {year} {2006})}\BibitemShut {NoStop}%
\bibitem [{\citenamefont {Chen}\ \emph {et~al.}(2010)\citenamefont {Chen},
  \citenamefont {Yang}, \citenamefont {Chen}, \citenamefont {Liu},
  \citenamefont {Wu}, \citenamefont {Zhan}, \citenamefont {Liang},\ and\
  \citenamefont {Wu}}]{Nanoplates}%
  \BibitemOpen
  \bibfield  {author} {\bibinfo {author} {\bibfnamefont {L.}~\bibnamefont
  {Chen}}, \bibinfo {author} {\bibfnamefont {X.}~\bibnamefont {Yang}}, \bibinfo
  {author} {\bibfnamefont {J.}~\bibnamefont {Chen}}, \bibinfo {author}
  {\bibfnamefont {J.}~\bibnamefont {Liu}}, \bibinfo {author} {\bibfnamefont
  {H.}~\bibnamefont {Wu}}, \bibinfo {author} {\bibfnamefont {H.}~\bibnamefont
  {Zhan}}, \bibinfo {author} {\bibfnamefont {C.}~\bibnamefont {Liang}},\ and\
  \bibinfo {author} {\bibfnamefont {M.}~\bibnamefont {Wu}},\ }\bibfield
  {title} {\bibinfo {title} {Continuous shape- and spectroscopy-tuning of
  hematite nanocrystals},\ }\href {https://doi.org/10.1021/ic100919a}
  {\bibfield  {journal} {\bibinfo  {journal} {Inorganic Chemistry}\ }\textbf
  {\bibinfo {volume} {49}},\ \bibinfo {pages} {8411} (\bibinfo {year}
  {2010})}\BibitemShut {NoStop}%
\bibitem [{\citenamefont {Liu}\ \emph {et~al.}(2012)\citenamefont {Liu},
  \citenamefont {Zhang}, \citenamefont {Wu}, \citenamefont {Yang},
  \citenamefont {Liu}, \citenamefont {Zhang}, \citenamefont {Wang},
  \citenamefont {Yao}, \citenamefont {Zhu},\ and\ \citenamefont
  {Zhao}}]{Bigcuboids}%
  \BibitemOpen
  \bibfield  {author} {\bibinfo {author} {\bibfnamefont {X.}~\bibnamefont
  {Liu}}, \bibinfo {author} {\bibfnamefont {J.}~\bibnamefont {Zhang}}, \bibinfo
  {author} {\bibfnamefont {S.}~\bibnamefont {Wu}}, \bibinfo {author}
  {\bibfnamefont {D.}~\bibnamefont {Yang}}, \bibinfo {author} {\bibfnamefont
  {P.}~\bibnamefont {Liu}}, \bibinfo {author} {\bibfnamefont {H.}~\bibnamefont
  {Zhang}}, \bibinfo {author} {\bibfnamefont {S.}~\bibnamefont {Wang}},
  \bibinfo {author} {\bibfnamefont {X.}~\bibnamefont {Yao}}, \bibinfo {author}
  {\bibfnamefont {G.}~\bibnamefont {Zhu}},\ and\ \bibinfo {author}
  {\bibfnamefont {H.}~\bibnamefont {Zhao}},\ }\bibfield  {title} {\bibinfo
  {title} {Single crystal $\alpha$-fe$_2$o$_3$ with exposed {104} facets for
  high performance gas sensor applications},\ }\href
  {https://doi.org/10.1039/C2RA20797D} {\bibfield  {journal} {\bibinfo
  {journal} {RSC Adv.}\ }\textbf {\bibinfo {volume} {2}},\ \bibinfo {pages}
  {6178} (\bibinfo {year} {2012})}\BibitemShut {NoStop}%
\bibitem [{\citenamefont {Young}(1995)}]{Young}%
  \BibitemOpen
  \bibfield  {author} {\bibinfo {author} {\bibfnamefont {R.}~\bibnamefont
  {Young}},\ }\href@noop {} {\emph {\bibinfo {title} {The Rietveld Method}}}\
  (\bibinfo  {publisher} {International Union of Crystallography. Oxford
  University Press},\ \bibinfo {year} {1995})\BibitemShut {NoStop}%
\bibitem [{\citenamefont {Hill}\ \emph {et~al.}(2008)\citenamefont {Hill},
  \citenamefont {Jiao}, \citenamefont {Bruce}, \citenamefont {Harrison},
  \citenamefont {Kockelmann},\ and\ \citenamefont {Ritter}}]{Hill}%
  \BibitemOpen
  \bibfield  {author} {\bibinfo {author} {\bibfnamefont {A.~H.}\ \bibnamefont
  {Hill}}, \bibinfo {author} {\bibfnamefont {F.}~\bibnamefont {Jiao}}, \bibinfo
  {author} {\bibfnamefont {P.~G.}\ \bibnamefont {Bruce}}, \bibinfo {author}
  {\bibfnamefont {A.}~\bibnamefont {Harrison}}, \bibinfo {author}
  {\bibfnamefont {W.}~\bibnamefont {Kockelmann}},\ and\ \bibinfo {author}
  {\bibfnamefont {C.}~\bibnamefont {Ritter}},\ }\bibfield  {title} {\bibinfo
  {title} {Neutron diffraction study of mesoporous and bulk hematite,
  $\alpha$-$fe_{2}o_{3}$},\ }\href {https://doi.org/10.1021/cm800009s}
  {\bibfield  {journal} {\bibinfo  {journal} {Chemistry of Materials}\ }\textbf
  {\bibinfo {volume} {20}},\ \bibinfo {pages} {4891} (\bibinfo {year}
  {2008})}\BibitemShut {NoStop}%
\bibitem [{\citenamefont {Joy}\ \emph {et~al.}(1998)\citenamefont {Joy},
  \citenamefont {Kumar},\ and\ \citenamefont {Date}}]{Joy}%
  \BibitemOpen
  \bibfield  {author} {\bibinfo {author} {\bibfnamefont {P.~A.}\ \bibnamefont
  {Joy}}, \bibinfo {author} {\bibfnamefont {P.~S.~A.}\ \bibnamefont {Kumar}},\
  and\ \bibinfo {author} {\bibfnamefont {S.~K.}\ \bibnamefont {Date}},\
  }\bibfield  {title} {\bibinfo {title} {The relationship between field-cooled
  and zero-field-cooled susceptibilities of some ordered magnetic systems},\
  }\href {http://stacks.iop.org/0953-8984/10/i=48/a=024} {\bibfield  {journal}
  {\bibinfo  {journal} {Journal of Physics: Condensed Matter}\ }\textbf
  {\bibinfo {volume} {10}},\ \bibinfo {pages} {11049} (\bibinfo {year}
  {1998})}\BibitemShut {NoStop}%
\bibitem [{\citenamefont {Blundell}(2001)}]{Blundell}%
  \BibitemOpen
  \bibfield  {author} {\bibinfo {author} {\bibfnamefont {S.}~\bibnamefont
  {Blundell}},\ }\href@noop {} {\emph {\bibinfo {title} {Magnetism in Condensed
  Matter,}}}\ (\bibinfo  {publisher} {Oxford University Press},\ \bibinfo
  {year} {2001})\BibitemShut {NoStop}%
\bibitem [{\citenamefont {Sandonis}\ \emph {et~al.}(1992)\citenamefont
  {Sandonis}, \citenamefont {Baruchel}, \citenamefont {Tanner}, \citenamefont
  {Fillion}, \citenamefont {Kvardakov},\ and\ \citenamefont
  {Podurets}}]{Sendonis}%
  \BibitemOpen
  \bibfield  {author} {\bibinfo {author} {\bibfnamefont {J.}~\bibnamefont
  {Sandonis}}, \bibinfo {author} {\bibfnamefont {J.}~\bibnamefont {Baruchel}},
  \bibinfo {author} {\bibfnamefont {B.}~\bibnamefont {Tanner}}, \bibinfo
  {author} {\bibfnamefont {G.}~\bibnamefont {Fillion}}, \bibinfo {author}
  {\bibfnamefont {V.}~\bibnamefont {Kvardakov}},\ and\ \bibinfo {author}
  {\bibfnamefont {K.}~\bibnamefont {Podurets}},\ }\href
  {https://doi.org/http://dx.doi.org/10.1016/0304-8853(92)90829-D} {\bibfield
  {journal} {\bibinfo  {journal} {Journal of Magnetism and Magnetic Materials}\
  }\textbf {\bibinfo {volume} {104}},\ \bibinfo {pages} {350 } (\bibinfo {year}
  {1992})}\BibitemShut {NoStop}%
\bibitem [{\citenamefont {Yanes}\ \emph {et~al.}(2013)\citenamefont {Yanes},
  \citenamefont {Jackson}, \citenamefont {Udvardi}, \citenamefont {Szunyogh},\
  and\ \citenamefont {Nowak}}]{Yanes}%
  \BibitemOpen
  \bibfield  {author} {\bibinfo {author} {\bibfnamefont {R.}~\bibnamefont
  {Yanes}}, \bibinfo {author} {\bibfnamefont {J.}~\bibnamefont {Jackson}},
  \bibinfo {author} {\bibfnamefont {L.}~\bibnamefont {Udvardi}}, \bibinfo
  {author} {\bibfnamefont {L.}~\bibnamefont {Szunyogh}},\ and\ \bibinfo
  {author} {\bibfnamefont {U.}~\bibnamefont {Nowak}},\ }\bibfield  {title}
  {\bibinfo {title} {Exchange bias driven by dzyaloshinskii-moriya
  interactions},\ }\href {https://doi.org/10.1103/PhysRevLett.111.217202}
  {\bibfield  {journal} {\bibinfo  {journal} {Phys. Rev. Lett.}\ }\textbf
  {\bibinfo {volume} {111}},\ \bibinfo {pages} {217202} (\bibinfo {year}
  {2013})}\BibitemShut {NoStop}%
\bibitem [{\citenamefont {Bae}\ \emph {et~al.}(2002)\citenamefont {Bae},
  \citenamefont {Judy}, \citenamefont {Chen},\ and\ \citenamefont
  {Egelhoff}}]{Bae}%
  \BibitemOpen
  \bibfield  {author} {\bibinfo {author} {\bibfnamefont {S.}~\bibnamefont
  {Bae}}, \bibinfo {author} {\bibfnamefont {J.~H.}\ \bibnamefont {Judy}},
  \bibinfo {author} {\bibfnamefont {P.~J.}\ \bibnamefont {Chen}},\ and\
  \bibinfo {author} {\bibfnamefont {W.~F.}\ \bibnamefont {Egelhoff}},\
  }\bibfield  {title} {\bibinfo {title} {Dependence of physical properties and
  giant magnetoresistance ratio on substrate position during rf sputtering of
  nio and $\alpha$-$fe_{2}o_{3}$ for bottom spin valves},\ }\href
  {https://doi.org/10.1063/1.1508161} {\bibfield  {journal} {\bibinfo
  {journal} {Applied Physics Letters}\ }\textbf {\bibinfo {volume} {81}},\
  \bibinfo {pages} {2208} (\bibinfo {year} {2002})}\BibitemShut {NoStop}%
\bibitem [{\citenamefont {Kawawake}\ \emph {et~al.}(1999)\citenamefont
  {Kawawake}, \citenamefont {Sugita}, \citenamefont {Satomi},\ and\
  \citenamefont {Sakakima}}]{Kawawake}%
  \BibitemOpen
  \bibfield  {author} {\bibinfo {author} {\bibfnamefont {Y.}~\bibnamefont
  {Kawawake}}, \bibinfo {author} {\bibfnamefont {Y.}~\bibnamefont {Sugita}},
  \bibinfo {author} {\bibfnamefont {M.}~\bibnamefont {Satomi}},\ and\ \bibinfo
  {author} {\bibfnamefont {H.}~\bibnamefont {Sakakima}},\ }\bibfield  {title}
  {\bibinfo {title} {Spin valves with a thin pinning layer of
  $\alpha$-fe$_2$o$_3$ or $\alpha$-$fe_{2}o_{3}$/nio},\ }\href
  {https://doi.org/10.1063/1.370079} {\bibfield  {journal} {\bibinfo  {journal}
  {Journal of Applied Physics}\ }\textbf {\bibinfo {volume} {85}},\ \bibinfo
  {pages} {5024} (\bibinfo {year} {1999})}\BibitemShut {NoStop}%
\bibitem [{\citenamefont {Rollmann}\ \emph {et~al.}(2004)\citenamefont
  {Rollmann}, \citenamefont {Rohrbach}, \citenamefont {Entel},\ and\
  \citenamefont {Hafner}}]{Rollmann}%
  \BibitemOpen
  \bibfield  {author} {\bibinfo {author} {\bibfnamefont {G.}~\bibnamefont
  {Rollmann}}, \bibinfo {author} {\bibfnamefont {A.}~\bibnamefont {Rohrbach}},
  \bibinfo {author} {\bibfnamefont {P.}~\bibnamefont {Entel}},\ and\ \bibinfo
  {author} {\bibfnamefont {J.}~\bibnamefont {Hafner}},\ }\bibfield  {title}
  {\bibinfo {title} {First-principles calculation of the structure and magnetic
  phases of hematite},\ }\href {https://doi.org/10.1103/PhysRevB.69.165107}
  {\bibfield  {journal} {\bibinfo  {journal} {Phys. Rev. B}\ }\textbf {\bibinfo
  {volume} {69}},\ \bibinfo {pages} {165107} (\bibinfo {year}
  {2004})}\BibitemShut {NoStop}%
\bibitem [{\citenamefont {Birss}(1964)}]{Birss}%
  \BibitemOpen
  \bibfield  {author} {\bibinfo {author} {\bibfnamefont {R.~R.}\ \bibnamefont
  {Birss}},\ }\href@noop {} {\emph {\bibinfo {title} {Symmetry and
  Magnetism}}}\ (\bibinfo  {publisher} {New York: North-Holland Pub. Co.},\
  \bibinfo {year} {1964})\BibitemShut {NoStop}%
\end{thebibliography}%
		
\end{document}